# Uncertainty Quantification and Global Sensitivity Analysis of Patient-Specific Coronary Hemodynamics Under Unsteady Flow Conditions

**Muhammad Usman[a,b], Nafis M. Arefin[a,b], Akil Narayan[c,d], Lucas H. Timmins[a,b,c,e,*]**

[a] Department of Biomedical Engineering, Texas A&M University, College Station, TX 77843, USA
[b] Institute of Biosciences and Technology, Texas A&M University, Houston, TX 77030, USA
[c] Scientific Computing and Imaging Institute, University of Utah, Salt Lake City, UT 84112, USA
[d] Department of Mathematics, University of Utah, Salt Lake City, UT 84112, USA
[e] Department of Biomedical Engineering, University of Utah, Salt Lake City, UT 84112, USA

***Corresponding author:**
Lucas H. Timmins, Ph.D.
Department of Biomedical Engineering
Texas A&M University
2121 W. Holcombe Blvd.
Houston, TX 77030, USA
lucas.timmins@tamu.edu
(979) 845-5532

## ABSTRACT

Patient-specific computational models of the coronary arteries demonstrate transformative potential for physiological assessment, identifying hemodynamic markers associated with disease progression, and optimizing treatment. Unfortunately, these models rarely account for input parameter uncertainty, limiting confidence in predictions and their clinical interpretation. Herein, we present a novel uncertainty quantification-finite element framework that integrates computationally efficient techniques to characterize and propagate uncertainty through patient-specific coronary simulations under unsteady flow conditions. Principal component analysis characterized variability in pulsatile coronary velocity waveforms, while uncertainty in blood pressure, viscosity, density, and heart rate is represented through probability distributions. Polynomial chaos expansion (PCE) enabled efficient uncertainty propagation and global sensitivity analysis across three patient-specific models, quantifying variability in time-averaged wall shear stress (WSS), oscillatory shear index, and cross-flow index. Fourth-order PCE emulators converged across all three patients, yielding analyses of 505 models per patient, with median coefficients of variation ranging from 0.15-0.20 for time-averaged WSS, 0.54-0.75 for oscillatory shear index, and 0.28-0.38 for cross-flow index, highlighting greater relative uncertainty in multidirectional WSS metrics. Global sensitivity analysis revealed that time-averaged WSS variability was governed by the individual effects of velocity and blood viscosity, with parameter interactions contributing <2% of total variance. In contrast, velocity waveform uncertainty dominated variability in the oscillatory shear index and cross-flow index, with higher-order parameter interactions accounting for 20.7-34.5% and 26.6-37.8% of total variance, respectively. These findings underscore the importance of incorporating physiological uncertainty into coronary simulations to promote confidence in predicted hemodynamics, particularly multidirectional WSS-derived metrics associated with atherosclerosis development and progression.

## 1. INTRODUCTION

Computational fluid dynamics (CFD) simulations are increasingly being integrated into clinical medicine, where they augment anatomic, physiologic, and functional data to inform disease diagnosis, prognosis, and treatment across the cardiovascular system [1]. This utility is particularly evident in the management of coronary artery disease (CAD), a leading cause of morbidity and mortality worldwide [2]. CFD-derived tools support multiple aspects of CAD assessment and management. Fractional flow reserve derived from coronary computed tomography angiography (FFR-CT) enables noninvasive assessment of lesion-specific ischemia to inform revascularization decisions in patients with CAD [3,4]. Wall shear stress (WSS)-based biomechanical metrics help identify lesions prone to progression and destabilization in acute coronary syndromes [5,6], and CFD simulations of percutaneous coronary intervention enable prediction of post-intervention physiological outcomes to inform procedural planning [7,8]. Central to each application is accurate, robust characterization of the patient-specific coronary hemodynamic environment.

In the epicardial coronary arteries, blood flow is inherently pulsatile, producing a time-varying hemodynamic environment characterized by metrics that integrate the WSS vector over the cardiac cycle, including time-averaged WSS (TAWSS), oscillatory shear index (OSI), and multidirectional measures such as cross-flow index (CFI) [9–11]. These complementary metrics capture aspects of the wall shear environment that WSS magnitude alone does not describe. In regions of complex geometry (e.g., bifurcations, curvature, stenoses), the WSS vector may change in magnitude, reverse direction, and develop components transverse to its temporal mean direction. This directional variability has important mechanobiological implications, as endothelial cells respond to both the magnitude and directionality of the applied WSS. For example, exposure to low, oscillatory, and multidirectional WSS promotes a pro-inflammatory, atheroprone endothelial phenotype that predisposes the vessel wall to lesion development and growth [12,13]. Consistent with these mechanistic observations, longitudinal clinical studies have demonstrated associations between these hemodynamic metrics and coronary atherosclerosis development and plaque progression [5,14–16], supporting their potential utility as prognostic markers of CAD.

Unfortunately, the computational models used to predict these hemodynamic metrics remain predominantly deterministic, whereby input parameters are assigned fixed values (often population means) and a single simulation provides the model output, neglecting the inherent variability (or uncertainty) in model inputs. In patient-specific coronary CFD models, uncertainty enters at multiple stages of the modeling pipeline, including vessel geometry reconstruction from finite-resolution clinical image data and the prescription of boundary conditions. These conditions may be derived from measurements subject to acquisition and physiological variability or from lumped-parameter models with uncertain parameters. Additional uncertainty arises from reliance on population-based literature values for blood material properties (i.e., viscosity, density). This input uncertainty, whether aleatoric or epistemic, is distinct from the numerical uncertainty addressed through formal verification [17,18], and propagates through the governing equations to the model output, leaving the uncertainty associated with the predicted hemodynamic metrics uncharacterized. As regulatory guidance increasingly emphasizes formal credibility assessment of computational models [19], establishing simulation frameworks that are robust to uncertainty is essential to support reliable model predictions and ensure models are accurate, reliable, and safe before clinical adoption.

Uncertainty quantification (UQ) provides a mathematical and statistical framework for characterizing variability in simulation outputs arising from uncertainty in model inputs, yielding output statistics, confidence measures, and parameter sensitivities [20,21]. Towards promoting confidence in blood flow simulations and model predictions, a range of UQ methods has been applied across vascular hemodynamic models, spanning Monte Carlo (MC) sampling, stochastic collocation, and surrogate-based approaches such as polynomial chaos expansion (PCE). Early studies applied stochastic collocation to parameterized carotid bifurcation and abdominal aortic aneurysm geometries to quantify uncertainty in predicted flow, pressure, and WSS [22], while subsequent studies extended UQ to clinically relevant coronary hemodynamic metrics such as CFD-derived FFR [23–26]. Applications now encompass patient-specific models of the coronary circulation [24,26], coronary artery bypass grafts [27], the aorta [28], and the pulmonary vasculature [29], with additional studies addressing uncertainty in the reconstructed vessel geometry itself [30,31]. Collectively, these studies demonstrate the utility and breadth of UQ for characterizing how input uncertainty propagates to hemodynamic predictions and identifying parameters that influence model outputs. However, implementing UQ in patient-specific CFD simulations requires careful consideration of which uncertain parameters to investigate, their probability distributions, the selection of appropriate UQ methods, and the scientific computing needs to obtain reliable uncertainty estimates. These challenges become particularly relevant for unsteady simulations, yet accounting for this variability is essential given the established role of time-dependent flow features and metrics in CAD development and progression.

Building on our previous investigations, the present study extends a coupled UQ-finite element (FE) framework that integrates the open-source UQ software `UncertainSCI` [32] with the `FEBio` software suite [33] to characterize uncertainty in patient-specific coronary artery hemodynamics under pulsatile (unsteady) flow conditions. This framework was previously applied to quantify how arterial material-parameter uncertainty propagates into variability in FE-predicted tissue stresses [34] and was extended to evaluate how hemodynamic parameter uncertainty affects WSS under steady flow conditions [35]. Although the steady flow investigation established the relative contributions of individual hemodynamic parameters and their interactions to WSS uncertainty, it did not account for temporal variability in coronary blood flow or its influence on the uncertainty of cardiac cycle-dependent hemodynamic metrics. Extending the framework to include pulsatile flow introduces additional methodological and computational challenges, including statistically characterizing and parameterizing time-varying inlet velocity waveforms, incorporating heart rate as an additional uncertain parameter, and increasing the computational demands of transient CFD simulations. Accordingly, the objective of this study was to extend and apply a PCE-based UQ framework incorporating principal component analysis (PCA) to represent velocity waveform variability and efficiently propagate uncertainty in physiological inputs through unsteady coronary CFD simulations. Specifically, we sought to quantify the magnitude and spatial distribution of uncertainty in TAWSS, OSI, and CFI, and to determine the relative contributions of individual uncertain parameters and their higher-order interactions to the uncertainty in these hemodynamic quantities of interest (QoIs).

## 2. METHODS

This study builds on an established UQ-FE framework developed in our previous investigations to characterize uncertainty in patient-specific computational models of coronary artery biomechanics [34,35]. As illustrated in Fig. 1, probability distributions were defined to describe uncertainty in patient-specific velocity and pressure measurements and physiological variability in heart rate, viscosity, and density. PCA was employed to characterize time-dependent variability in the measured velocity waveforms. Parameter sets were generated within `UncertainSCI`[1] [32,36] and incorporated into a forward FE framework to perform unsteady CFD simulations in `FEBio`[2] [33]. WSS-derived quantities were calculated from the CFD outputs and used to construct PCE emulators for uncertainty quantification and global sensitivity analysis.

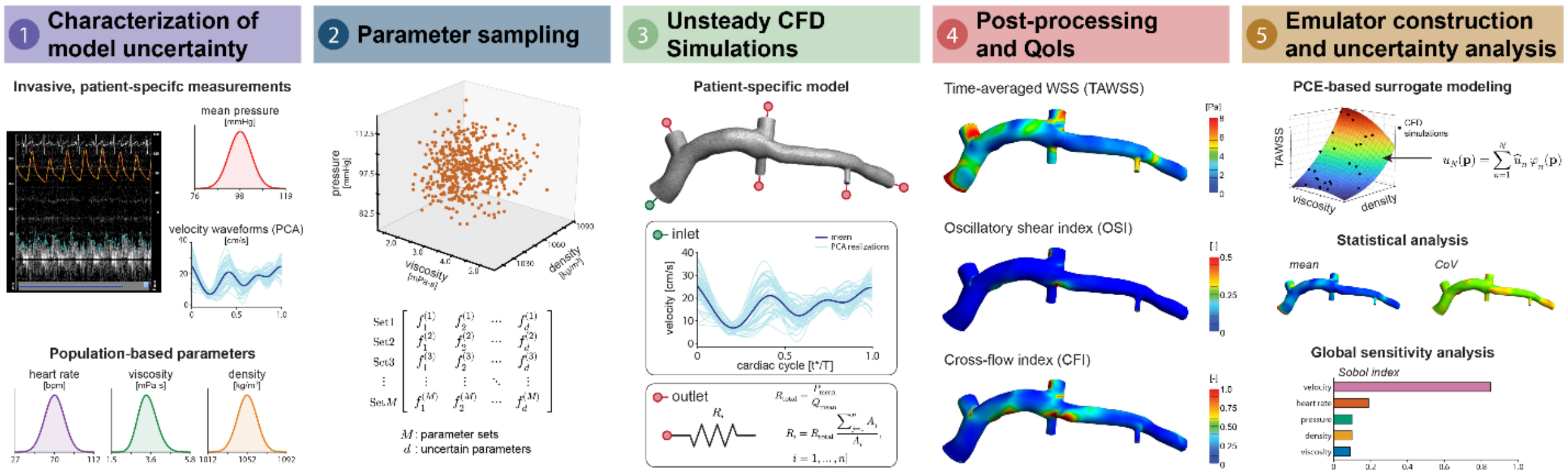

**Fig. 1.** Schematic overview of the UQ-FE framework in `UncertainSCI` [32] and `FEBio` [33]. The framework propagates uncertainty in patient-specific hemodynamic measurements and physiological parameters through unsteady coronary computational fluid dynamics (CFD) simulations to characterize variability and identify dominant sources of uncertainty in predicted hemodynamic QoIs.

## 2.1 Invasive Coronary Imaging and Hemodynamic Measurements

[1] Publicly hosted - https://github.com/SCIInstitute/UncertainSCI
[2] Publicly hosted - https://febio.org/

Clinical imaging and hemodynamic data were obtained from three patients enrolled in a previous prospective investigation of non-obstructive coronary artery disease progression (ClinicalTrials.gov identifier: NCT00576576). Briefly, patients referred for invasive coronary evaluation following an abnormal cardiac stress test underwent coronary angiography and virtual histology intravascular ultrasound of the left main and left anterior descending coronary arteries. Imaging data were combined to reconstruct patient-specific three-dimensional (3D) coronary geometries. Invasive velocity and pressure measurements were acquired in the left main (LM) and left anterior descending (LAD) coronary arteries over consecutive cardiac cycles (≥6). Additional details regarding patient characteristics, clinical data acquisition, and imaging protocols have been reported previously [5,37]. The Emory University Institutional Review Board approved the study protocol, and all patients provided written informed consent before participation.

**2.2 Input hemodynamic parameters and uncertainty distributions**

Five uncertain hemodynamic inputs were considered to characterize the influence of physiological and measurement variability on predicted coronary hemodynamics: velocity [$v(t)$], pressure ($P$), viscosity ($\mu$), density ($\rho$), and heart rate ($HR$). Patient-specific velocity and pressure measurements acquired as described in Section 2.1 were used to characterize their uncertainty distributions, whereas probability distributions for heart rate, viscosity, and density were derived from population-based data reported in the literature.

To characterize uncertainty in patient-specific velocity waveforms, PCA was employed to identify the dominant modes of waveform variability. Velocity data across individual cardiac cycles were digitized, concatenated into a continuous signal, and passed through a fourth-order Butterworth low-pass filter (normalized cutoff frequency = 0.3) with zero-phase forward-backward filtering (Fig. 2). The filtered signal was subsequently segmented into 6 or 7 individual cardiac cycles, depending on the number of cycles available for each patient. To account for differences in cardiac cycle duration, each waveform was mapped onto a normalized cardiac cycle and resampled at 300 equally spaced time points using cubic interpolation. This procedure established a consistent temporal representation of the measured waveforms, enabling direct comparison of cycle-to-cycle variations.

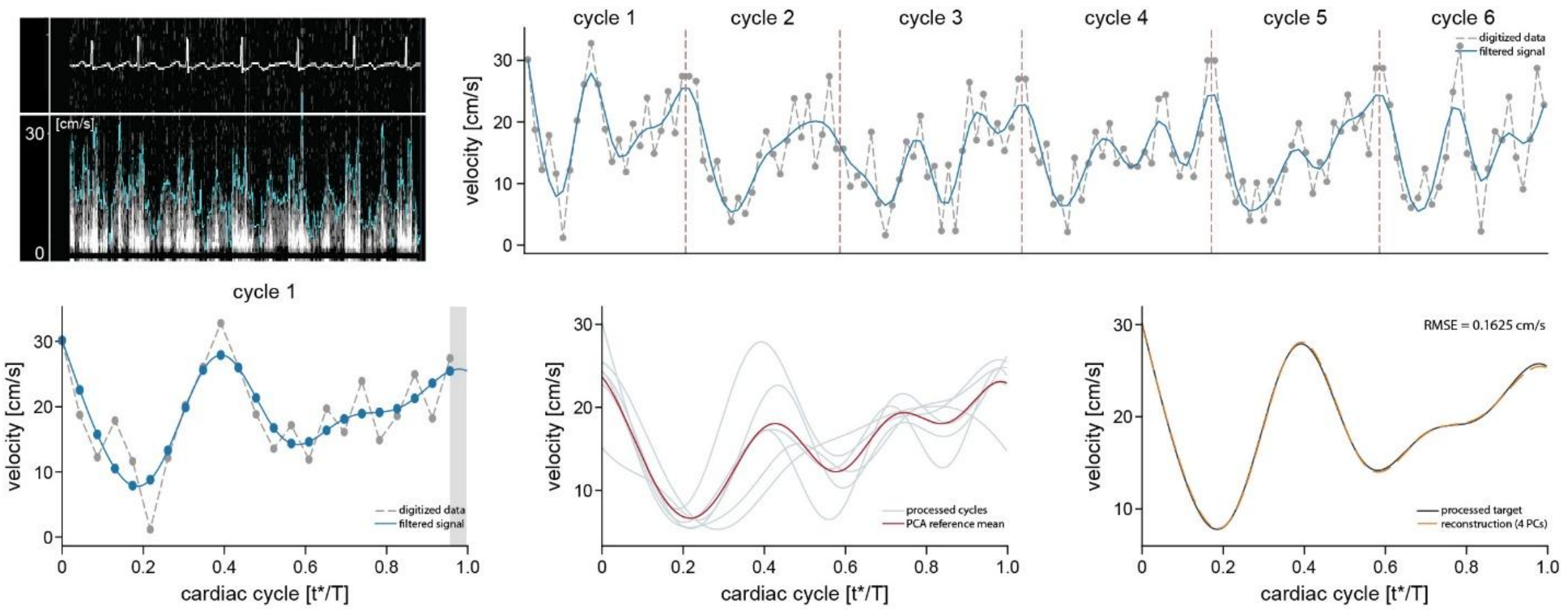

**Fig. 2.** Processing of patient-specific coronary velocity waveforms for principal component analysis (PCA). Invasive Doppler ultrasound measurements were digitized to obtain velocity data across consecutive cardiac cycles and filtered (top). The filtered signal was segmented into individual cycles, and cycle-to-cycle variability was characterized to establish a reference mean for PCA (bottom). A representative reconstruction of the processed waveform using four principal components, which closely agreed with the original waveform (bottom right). Note the shaded region in the filtered individual signal indicates the final 10 time points adjusted using interpolation to ensure a smooth transition between consecutive waveforms in the simulations (bottom left).

PCA was then performed separately across each patient's data using the normalized velocity waveforms to identify the dominant modes of cycle-to-cycle variability. Each waveform, $v_j(t)$, was represented as a linear combination of the mean velocity and principal components weighted by their corresponding scores:

$$v_j(t) = \bar{v}(t) + \sum_{k=1}^{K} a_{jk}\, \phi_k(t),$$

where $\bar{v}\,(t)$ is the mean velocity waveform, $\phi_k(t)$ is the $k$-th principal component, $a_{jk}$is its corresponding score for waveform $j$, and $K$ is the number of retained principal components. The cumulative explained variance was calculated, and the minimum number of principal components needed to explain approximately 95% of the total waveform variance was retained, balancing the need to capture the dominant sources of velocity waveform uncertainty with limiting the dimensionality of the uncertain parameter space and associated computational costs.

Probability distributions were defined within `UncertainSCI` to characterize variability in the uncertain hemodynamic parameters (Table 1). Patient-specific pressure distributions were characterized using the mean and standard deviation of cycle-averaged pressures from invasive measurements across multiple cardiac cycles. The heart rate distribution was derived from measurements obtained from the larger patient cohort of the clinical investigation [5], from which the three patients evaluated in the present study were selected, whereas viscosity and density distributions were based on population-based data reported in the literature [38,39]. For velocity, the PCA representation was incorporated into the framework by treating the retained principal component scores as uncertain parameters. Probability distributions were assigned to the scores to characterize the variability in waveform shape (Fig. 2). Normal distributions were assigned to most principal component scores, whereas beta distributions were used for selected components, with bounds prescribed to prevent negative velocities in the reconstructed waveforms (Table 3). The resulting distributions were evaluated across the generated parameter samples to confirm that all reconstructed velocity waveforms remained positive while capturing the dominant patterns of variability in the observed patient-specific measurements.

**Table 1.** Uncertain input parameters and associated probability distributions for pressure, viscosity, density, and heart rate. Patient-specific pressure distributions were derived from invasive measurements, whereas literature-derived values include viscosity [38], density [39], and heart rate [5]. Values are reported as mean ± standard deviation.

| Parameter | Distribution | Patient 1 | Patient 2 | Patient 3 |
|---|---|---|---|---|
| *Patient-specific parameter* | | | | |
| Pressure, $P$ | normal | 13,004.1 ± 722.4 Pa (97.54 ± 5.42 mmHg) | 12,591.9 ± 1,584.6 Pa (94.45 ± 11.89 mmHg) | 21,790.2 ± 2,093.4 Pa (163.44 ± 15.70 mmHg) |
| *Population-based parameters* | | | | |
| Viscosity, $\mu$ [mPa·s] | gamma | 3.48 ± 0.51 | | |
| Density, $\rho$ [kg/m³] | normal | 1,051.50 ± 10 | | |
| Heart rate, HR [bpm] | normal | 69.65 ± 10.57 | | |

## 2.3 Parameter Set Generation

The probability distributions were provided as inputs to `UncertainSCI` [32], which employs weighted approximate Fekete points (WAFP) for efficient parameter sampling. Briefly, WAFP generates a large candidate set of parameter combinations sampled from the provided probability distributions and subsequently employs a greedy, weighted *D*-optimal selection procedure to identify a subset that approximately maximizes the determinant of the weighted polynomial design matrix. The selected parameter set, $\mathcal{S}_{WAFP}$, is determined according to the following criterion:

$$\mathcal{S}_{WAFP} \approx \underset{\mathcal{S}\subset\mathcal{C},|\mathcal{S}|=M}{\arg\max} \det(\mathrm{V}_w^T \mathrm{V}_w),$$

where $\mathcal{S}_{WAFP}$ is the selected parameter set, $\mathcal{C}$ is the candidate set, $M$ is the number of selected samples, and $V_w$ is the weighted polynomial design matrix corresponding to the selected samples. WAFP has demonstrated superior numerical performance compared with conventional randomized and deterministic sampling approaches [40,41]. Moreover, our previous coronary hemodynamics UQ investigation demonstrated that WAFP-based PCE achieved converged uncertainty and sensitivity estimates using more than 10,000-fold fewer model evaluations than Monte Carlo sampling in an analytical flow benchmark [35].

Particularly important for computationally expensive, 3D patient-specific CFD simulations, this approach enables efficient sampling of the uncertain parameter space while substantially reducing the number of model evaluations required to construct accurate PCE surrogate models. The number of polynomial basis elements at each order was determined by:

$$N_p = \binom{d+p}{p},$$

where $N_p$ is the number of basis elements, $d$ is the dimension of the uncertain parameter space, and $p$ is the polynomial order. Following the additive oversampling strategy demonstrated by Burk et al. [41], the number of parameter sampling points, $N_s$, was defined as $N_s = N_p + f_S^A$, where $f_S^A$ = 10. These additional sampling points ($f_S^A$) were included beyond $N_p$ at each polynomial order to improve the stability and accuracy of the PCE surrogate construction without substantially increasing computational cost.

## 2.4 Patient-specific CFD Model Development and Simulation

Three patient-specific coronary geometries were reconstructed and meshed following procedures established in our previous investigations [35]. Briefly, 3D geometries of the left main and left anterior descending coronary arteries, and their major branches, were reconstructed from biplane angiography and electrocardiogram-gated virtual histology intravascular ultrasound (IVUS) imaging data [5,37,42]. Cross-sectional lumen boundaries were positioned perpendicular to the IVUS catheter, as determined from angiography, and combined to generate the coronary lumen surface. The geometries were discretized using `SimVascular` [43], following the meshing parameters demonstrated to yield mesh-converged solutions [35], including a maximum element size of 0.22 mm, 4 boundary layers, a portion of edge size of 0.5, and a boundary-layer decreasing ratio of 0.8. The resulting computational meshes contained approximately 1-2 million linear tetrahedral elements per patient-specific model.

Unsteady CFD simulations were performed using the CFD solver in `FEBio` [44]. Time-dependent velocity waveforms were applied as a series of plug profiles at the inlet, and resistance boundary conditions were prescribed at all outlet surfaces (Fig. 1). To ensure continuity in the prescribed inlet velocity between consecutive cardiac cycles, the final 10 time points (out of 300) of each preceding cycle were adjusted using cubic-spline interpolation to transition smoothly to the initial velocity of the subsequent cycle (Fig. 2). This procedure eliminated discontinuities arising from differences between the initial and final velocities of the generated waveforms, thereby preventing abrupt changes in velocity that could adversely affect numerical convergence. Importantly, the final cardiac cycle, from which hemodynamic quantities were evaluated, remained unchanged. Outlet resistance values were calculated from area scaling of the total resistance ($R_{tot}$) of the system, which was calculated as,

$$R_{tot} = \frac{P}{Q},$$

where $P$ is the sampled mean pressure and $Q$ is the time-averaged volumetric flow rate calculated from the prescribed inlet velocity waveform. The total resistance was subsequently distributed among the outlet branches according to their respective cross-sectional areas. A rigid wall assumption was employed, and a no-slip boundary condition was applied at the vessel wall.

Broyden's quasi-Newton method, in combination with the `mkl_dss` linear solver and automatic time stepping, was employed to facilitate numerical convergence [45]. Numerical solution settings were consistent with our previous investigation [35], except that the time-step size and total simulation duration were adjusted according to the sampled heart rate. Cardiac cycle periodicity was assessed by comparing element-wise TAWSS distributions over six consecutive cycles. In a representative patient-specific simulation, the relative difference in TAWSS between cycles 2 and 3 was ~0.9%, ~0.8%, and ~0.9% for the median, first quartile, and third quartile, respectively. Temporal sampling convergence was assessed by increasing the number of prescribed output time points, termed *must-point* in `FEBio` [33], from 30 to 60 per cardiac cycle, resulting in a ~0.9% relative difference in median TAWSS, negligible change in the first quartile, and a ~1.1% difference in the third quartile. Accordingly, each CFD simulation was performed over three cardiac cycles, with WSS magnitude and Cauchy stress tensor components exported at 30 equally spaced time points during the final cardiac cycle for postprocessing. Batch processing of the CFD simulations was executed on the `GRACE` (940 nodes, featuring Intel Xeon Gold 6248R processors at 3.0 GHz with 48 cores and 384 GB RAM per node) and

`FASTER` (180 nodes, featuring Intel Xeon Platinum 8352Y processors at 2.2 GHz with 64 cores and 256 GB RAM per node) clusters at the High Performance Research and Computing facility at Texas A&M University.

### 2.5 Hemodynamic Quantities of Interest

Three hemodynamic QoIs were evaluated based on their established association with coronary artery disease development and progression, as well as atherosclerosis more broadly: TAWSS [5,6,15,42], OSI [9,10], and CFI [11,14], representing the magnitude, directional oscillation, and multidirectionality of the WSS vector, respectively. TAWSS was calculated at each element by temporally averaging the instantaneous WSS vector magnitude, obtained directly from `FEBio`, over the cardiac cycle:

$$\text{TAWSS} = \frac{1}{T}\int_0^T \|\boldsymbol{\tau}_w(t)\|\, dt,$$

where $\boldsymbol{\tau}_w(t)$ is the instantaneous wall shear stress vector at time $t$ and $T$ is the duration of the cardiac cycle. OSI was calculated at each element to quantify the directional oscillation of WSS over the cardiac cycle:

$$\text{OSI} = \frac{1}{2}\left[1 - \frac{\left\|\int_0^T \boldsymbol{\tau}_w(t)dt\right\|}{\int_0^T \|\boldsymbol{\tau}_w(t)\|dt}\right],$$

OSI is a dimensionless quantity ranging from 0 to 0.5, where values approaching 0 indicate predominantly unidirectional WSS and values approaching 0.5 indicate pronounced directional oscillation. CFI quantifies the directionality of the WSS vector relative to the temporal mean direction, independent of WSS magnitude, and is defined as:

$$\mathrm{CFI} = \frac{1}{T}\int_0^T \left| \frac{\boldsymbol{\tau}_w(t)}{|\boldsymbol{\tau}_w(t)|} \cdot \left( \mathbf{n} \times \frac{\int_0^T \boldsymbol{\tau}_w\,(t)\,dt}{\left|\int_0^T \boldsymbol{\tau}_w\,(t) dt\right|} \right) \right| dt,$$

where **n** is the unit vector normal to the vessel wall. CFI is a dimensionless quantity ranging from 0 to 1, with higher values indicating a greater contribution of transverse WSS vector components relative to the temporal mean vector direction.

### 2.6 Polynomial Chaos Expansion-Based Uncertainty Quantification and Sensitivity Analysis

Building on our previously established UQ frameworks and related investigations [34,35], forward UQ and global sensitivity analysis were performed using the open-source Python framework `UncertainSCI` [32]. Post-processed CFD outputs served as inputs to `UncertainSCI` to construct PCE emulators relating uncertain input parameters to predicted hemodynamic QoIs. The PCE emulator approximated the CFD model response as a linear combination of multivariate orthogonal polynomial functions, with the number of polynomial terms specified by the user, expressed as:

$$u(\boldsymbol{p}) \approx u_N(\boldsymbol{p}) = \sum_{n=1}^{N} \hat{u}_n \phi_n(\boldsymbol{p}),$$

where $u(\boldsymbol{p})$ represents the CFD model output for a given set of uncertain input parameters $\boldsymbol{p}$, $u_N(\boldsymbol{p})$ is the corresponding PCE approximation, $N$ is the number of polynomial functions, $\hat{u}_n$ represents the polynomial coefficients that are computed from the simulation data, and $\phi_n(\boldsymbol{p})$ are the orthogonal polynomial functions in $\boldsymbol{p}$. Once constructed, the resulting emulator provides a computationally efficient surrogate for the patient-specific CFD simulations and enables direct calculation of uncertainty statistics and global sensitivity indices without requiring additional simulations.

To assess the quality of the constructed emulators, the relative error ($\varepsilon_\delta$) was quantified between the emulator approximations and the predicted QoIs as:

$$\varepsilon_\delta = \frac{1}{N}\sum_{i=1}^{N} \frac{\left\|A\hat{x} - \vec{b}\right\|_2}{\left\|\vec{b}\right\|_2},$$

where $A\hat{x}$ is the approximation, $\vec{b}$ is the FE model-predicted QoI, $N$ is the number of elements, and $\|\cdot\|_2$ denotes the 2-norm of the vector. PCE convergence was evaluated across polynomial orders $p = \{1, 2, 3, 4\}$ for each patient to determine the order required to sufficiently approximate the QoIs while minimizing computational cost. For Patient 1, three independent sampling runs were performed at each polynomial order to assess the magnitude and variability of relative error across independent PCE constructions. A single sampling run was performed at each polynomial order for Patients 2 and 3 to reduce computational cost.

Statistical measures, including the mean, standard deviation, and coefficient of variation (CoV), were determined directly from the constructed PCE emulators to quantify how input parameter uncertainty propagates to uncertainty in the predicted hemodynamic QoIs. These measures were calculated at each vessel wall element in the computational model to characterize the spatial distribution of uncertainty in TAWSS, OSI, and CFI. Global sensitivity analyses were subsequently performed using Sobol's indices [46] to quantify the relative contributions of uncertain input parameters to variability in predicted QoIs. First-order ($S_i$) and total-order ($S_i^T$) Sobol sensitivity indices were calculated directly from the polynomial chaos expansion (PCE) coefficients following Burk et al. (2020), as:

$$S_i = \frac{\sum_{\boldsymbol{\alpha} \in \mathcal{A}_i} c_{\boldsymbol{\alpha}}^2}{\sum_{\boldsymbol{\alpha} \neq \mathbf{0}} c_{\boldsymbol{\alpha}}^2},$$

and

$$S_i^T = \frac{\sum_{\boldsymbol{\alpha} \in \mathcal{A}_i^T} c_{\boldsymbol{\alpha}}^2}{\sum_{\boldsymbol{\alpha} \neq \mathbf{0}} c_{\boldsymbol{\alpha}}^2},$$

where $c_\alpha$ represents the PCE coefficient corresponding to the multi-index $\alpha$. The index set $\mathcal{A}_i$ contains polynomial terms involving only the $i^{\text{th}}$ uncertain input, whereas $\mathcal{A}_i^T$ includes all terms involving that input, including interactions with other uncertain parameters. First-order Sobol indices quantify the independent contribution of each uncertain input to the variance of the predicted QoI, allowing parameters to be ranked according to their relative influence. Total-order Sobol indices additionally account for interactions with other uncertain inputs, providing a measure of the overall contribution of each parameter to output variability. Since uncertainty in the inlet velocity waveform was represented by principal component scores, these components were grouped to quantify the overall contribution of velocity uncertainty to the predicted QoIs. Grouped total-, first-, and higher-order Sobol indices were calculated for the velocity parameter group, allowing direct comparison of its contribution with that of the other uncertain parameters. Table 2 summarizes the UQ and FE setup for each patient-specific model, including the parameter sets per PCE order, sampling runs, and total CFD simulations.

**Table 2.** Summary of UQ and FE setup for each patient-specific model.

| Parameter | Patient 1 | Patient 2 | Patient 3 |
|---|---|---|---|
| Measured cardiac cycles | 7 | 6 | 6 |
| Mesh size (elements) | ~2.03M | ~1.07M | ~1.01M |
| Uncertain dimensions | | 8 | |
| Parameter groups | | 5 – $v$ (grouped from 4 PCs), $P$, $\mu$, $\rho$, HR | |
| PCE orders evaluated | | 1–4 | |
| Parameter sets per PCE order (1–4) | | 19 / 55 / 175 / 505 | |
| Independent sampling runs | 3 | 1 | 1 |
| Total CFD simulations | 2,262 | 754 | 754 |

The PCE convergence criteria were defined as a mean relative error of <2.5% and first-order Sobol index value changes of <0.06 between increasing polynomial orders (i.e., index stability). Data are reported as mean ± standard deviation for normally distributed data and as median (Q1, Q3) for non-normally distributed variables.

## 3. RESULTS

### 3.1 Principal Component Analysis and Parameter Sampling

PCA showed that four principal components captured 96.5%, 93.8%, and 99.6% of the variability in the coronary velocity waveforms for Patients 1, 2, and 3, respectively (Fig. 3). For Patient 1, the cumulative explained variance increased from 59.6% for the first principal component to 81.2%, 88.9%, and 96.5% with the inclusion of the second, third, and fourth components, respectively. Similar trends were observed for Patients 2 and 3, although the cumulative explained variance increased more rapidly for Patient 3, with three components sufficient to explain approximately 95% of the waveform variability. To maintain consistency across patients, four principal components were retained for all subsequent UQ analyses. Together with pressure, viscosity, density, and heart rate, this resulted in eight uncertain dimensions for each patient.

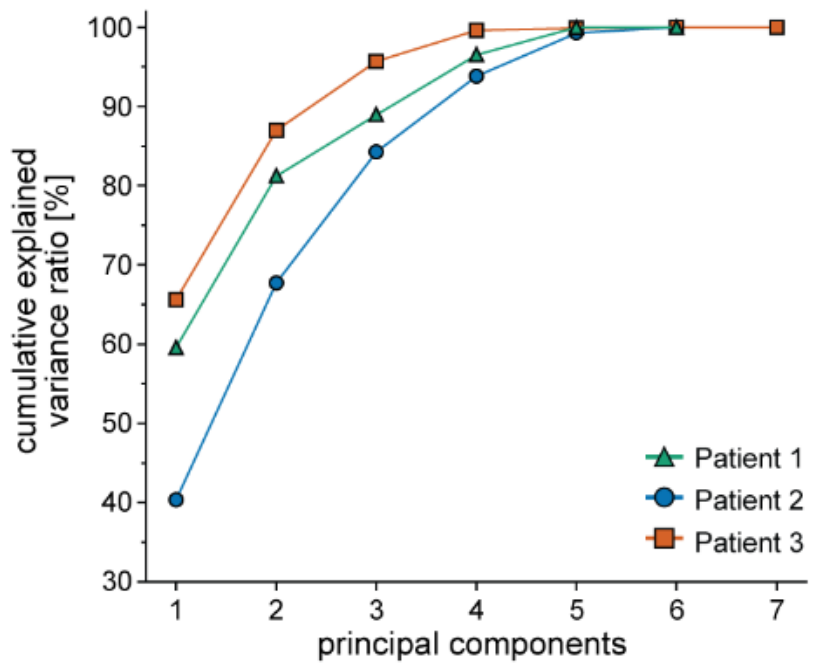

**Fig. 3.** Cumulative explained variance ratio as a function of the number of principal components for the patient-specific coronary velocity waveforms.

Table 3 summarizes the probability distributions associated with the four principal components, including adjustments to selected distributions to ensure positive velocities in the reconstructed waveforms (e.g., beta distribution for principal component 1). These distributions were sampled using `UncertainSCI` to characterize variability in the patient-specific coronary velocity waveforms. Representative velocity waveforms generated from the second-order PCE parameter sets are shown for each patient at PCE-order 2 in Fig. 4, illustrating the variability in waveform magnitude and temporal characteristics incorporated into the UQ analysis.

**Table 3.** Probability distributions and corresponding parameters of the principal component scores used to characterize uncertainty in the patient-specific coronary velocity waveforms. Normal distributions are described by their mean ($\mu$) and standard deviation ($\sigma$), whereas beta distributions are described by their shape parameters (i.e., $[\alpha, \beta]$).

| Principal Component | Distribution | Patient 1 | Patient 2 | Patient 3 |
|---|---|---|---|---|
| 1 | beta | [4.0, 4.0] | [3.8, 4.1] | [2.32, 4.7] |
| 2 | normal/beta | $\mu = 0, \sigma = 27.9$ | [2.2, 4.6] | [2.6, 4.7] |
| 3 | normal | $\mu = 0, \sigma = 16.7$ | $\mu = 0, \sigma = 40.5$ | $\mu = 0, \sigma = 33.9$ |
| 4 | normal | $\mu = 0, \sigma = 16.5$ | $\mu = 0, \sigma = 30.8$ | $\mu = 0, \sigma = 22.7$ |

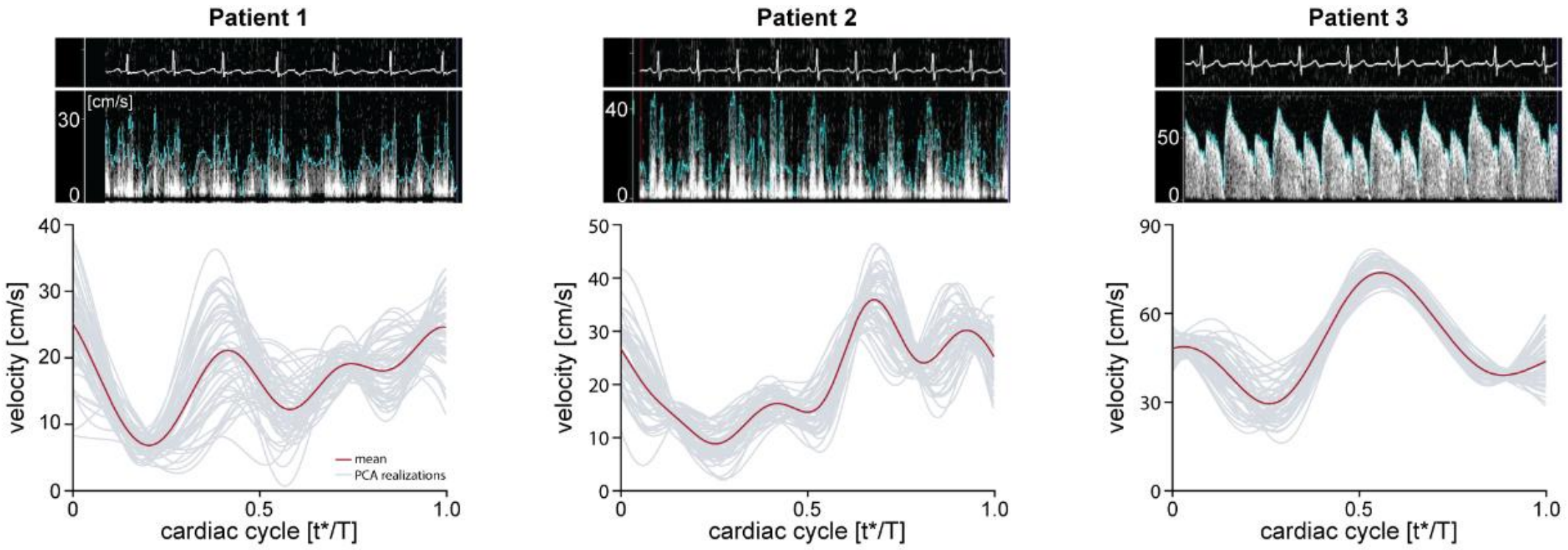

**Fig. 4.** Patient-specific coronary velocity waveforms generated from 55 parameter sets sampled for second-order PCE. For each patient, the original Doppler velocity and EKG tracings (top) and corresponding mean velocity waveform and PCA-based realizations over a normalized cardiac cycle (bottom).

## 3.2 PCE Convergence

The number of parameter sets generated by `UncertainSCI` ranged from 19 (order 1) to 55 (order 2), 175 (order 3), and 505 (order 4). Convergence results for Patient 1, based on three independent sampling runs at each polynomial order, are presented in Fig. 5. Spatially averaged residual errors decreased progressively with increasing PCE order for all QoIs. From first- to fourth-order PCE, errors decreased from 1.46 ± 0.17% to 0.04 ± 0.002% for TAWSS, 23.4 ± 4.0% to 2.19 ± 0.09% for OSI, and 12.3 ± 2.29% to 1.47 ± 0.11% for CFI. First-order Sobol indices also became increasingly stable with increasing polynomial order. The mean absolute changes in first-order Sobol indices across the five parameter groups between third- and fourth-order PCE were 0.0046, 0.0071, and 0.0312 for TAWSS, OSI, and CFI, respectively. At fourth order, the standard deviations across the three independent runs of the mean first-order Sobol indices (averaged across the five parameter groups) were 0.0007, 0.0107, and 0.0056, respectively.

Similar convergence trends were observed for Patients 2 and 3, with residual errors decreasing progressively from first- to fourth- order PCE across all QoIs (Figs. S2 and S3). At fourth order, residual errors were 0.024%, 1.42%, and 1.04% for TAWSS, OSI, and CFI, respectively, in Patient 2, compared with 0.025%, 1.27%, and 0.76% in Patient 3. First-order Sobol indices also became increasingly stable with polynomial order. The mean absolute changes in first-order Sobol indices across the five parameter groups between third- and fourth-order PCE were 0.0016, 0.0102, and 0.0169 for TAWSS, OSI, and CFI, respectively, in Patient 2, compared with 0.0010, 0.0109, and 0.0160 in Patient 3. Thus, fourth-order PCE satisfied the specified convergence criteria, with residual errors below 2.5% and changes in all individual Sobol indices below 0.06 between third- and fourth-order PCE, supporting its selection for subsequent UQ analyses. Example parameter distributions at order 4 for Patient 1 are shown in Fig. S1.

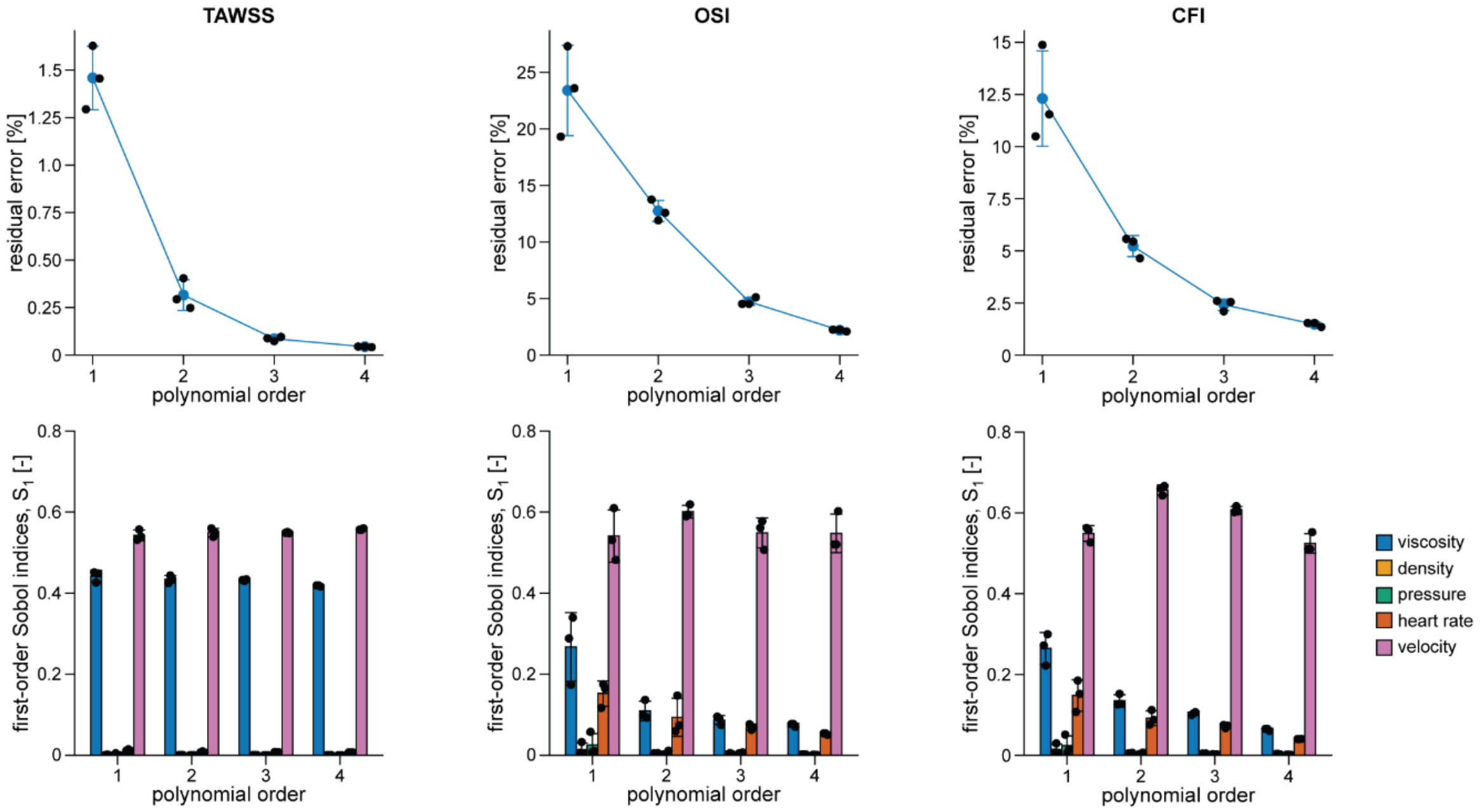

**Fig. 5.** PCE convergence analysis for Patient 1 across PCE orders 1-4. Spatially averaged residual errors (top row) and first-order Sobol indices (bottom row) are presented for time-averaged WSS (TAWSS), oscillatory shear index

(OSI), and cross-flow index (CFI). Results are shown for three independent PCE runs at each order, with individual runs indicated by black markers. Data are reported as mean ± standard deviation.

The accuracy of the fourth-order PCE emulator was further assessed through analysis of instantaneous WSS magnitudes at 30 time points throughout the final cardiac cycle for each of the 505 simulations in Patient 1 (Fig. 6). Spatially averaged residual errors remained below 0.5% throughout the cycle, ranging from 0.06% to 0.47%. Notably, residual error was below 0.24% at the cardiac phase corresponding to maximum WSS uncertainty (i.e., maximum WSS standard deviation; t*/T = 0.379). Time-resolved total Sobol indices demonstrated that velocity uncertainty dominated instantaneous WSS variability throughout most of the cardiac cycle, with viscosity becoming the principal contributor during two brief intervals. Contributions from density, pressure, and heart rate remained negligible.

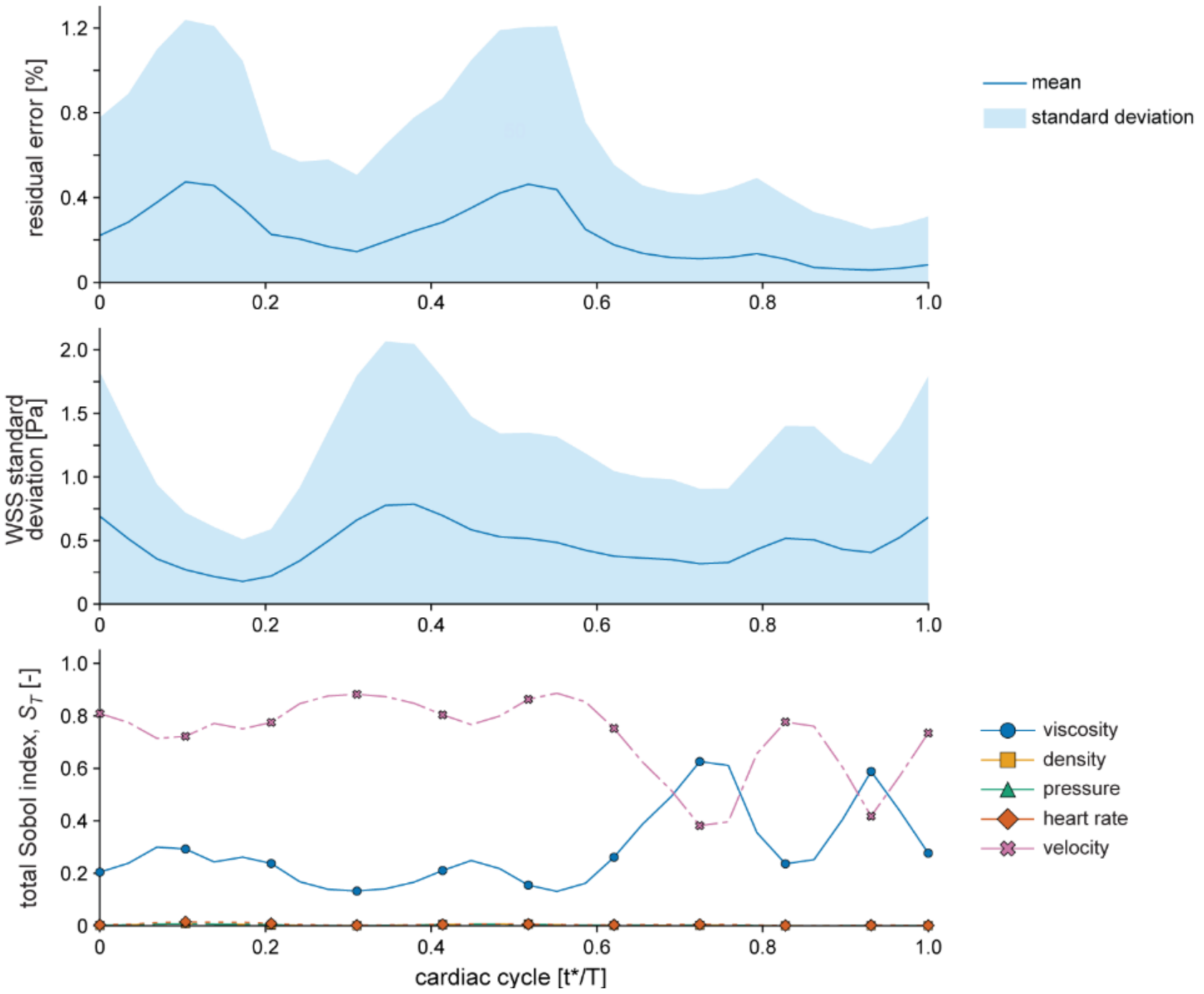

**Fig. 6.** Time-resolved uncertainty and sensitivity analysis of instantaneous wall shear stress (WSS) throughout the cardiac cycle for Patient 1 using fourth-order PCE. Spatially averaged PCE residual error (top row), standard deviation of WSS (middle row), and total Sobol sensitivity indices for the uncertain parameters (bottom row). Data are presented as a function of normalized cardiac cycle (t*/T).

### 3.3 Uncertainty in Hemodynamic Quantities of Interest

Fig. 7 presents the spatial distributions of mean TAWSS, OSI, and CFI across the three patient-specific models, calculated using fourth-order PCE emulators constructed from 505 CFD simulations per patient. All three QoIs exhibited spatial heterogeneity, with localized regions of low TAWSS, elevated OSI, and elevated CFI occurring near vessel branches, regions of curvature, and abrupt geometric changes. Median (Q1, Q3) values of TAWSS were 1.048 Pa (0.673, 1.785), 2.584 Pa (1.647, 4.361), and 1.953 Pa (0.771, 3.527) for Patients 1, 2, and 3, respectively. Corresponding values for OSI were 0.00174 (0.00037, 0.00597), 0.00103 (0.00025, 0.00413), and 0.00043 (0.00011, 0.00147), whereas values for CFI were 0.0670 (0.0303, 0.1268), 0.0579 (0.0272, 0.1156), and 0.0346 (0.0176, 0.0633), respectively. Despite differences in the magnitude and spatial distribution of the QoIs among patients, regions of elevated OSI and CFI generally coincided in areas where TAWSS was reduced.

To quantify the relative magnitude of uncertainty in the predicted hemodynamic QoIs, spatial distributions of the CoV (i.e., standard deviation/mean) were evaluated across all three patients (Fig. 8). TAWSS exhibited relatively uniform CoV across the vessels, whereas OSI and CFI displayed greater spatial heterogeneity, particularly near vessel branches and regions of curvature. Across all patients, OSI consistently exhibited the greatest relative uncertainty, followed by CFI and TAWSS. Spatial median (Q1, Q3) CoV values for OSI were 0.75 (0.56, 1.23), 0.62 (0.44, 1.02), and 0.54 (0.43, 0.75) for Patients 1, 2, and 3, respectively. In comparison, median CoV values for TAWSS and CFI ranged from 0.15 to 0.20 and 0.28 to 0.38, respectively, across the three patients. Expressed relative to their respective mean values, the standard deviation of the predicted QoIs ranged from approximately 15-20% for TAWSS, 54-75% for OSI, and 28-38% for CFI. Notably, OSI exhibited a CoV greater than 0.4 across at least 75% of the lumen surface in all three patients, with CoV values exceeding 1.0 across a minimum of 25% of the lumen surface in Patients 1 and 2. Even in Patient 3, which exhibited the lowest median OSI uncertainty, CoV values reached or exceeded 0.75 across at least 25% of the lumen surface.

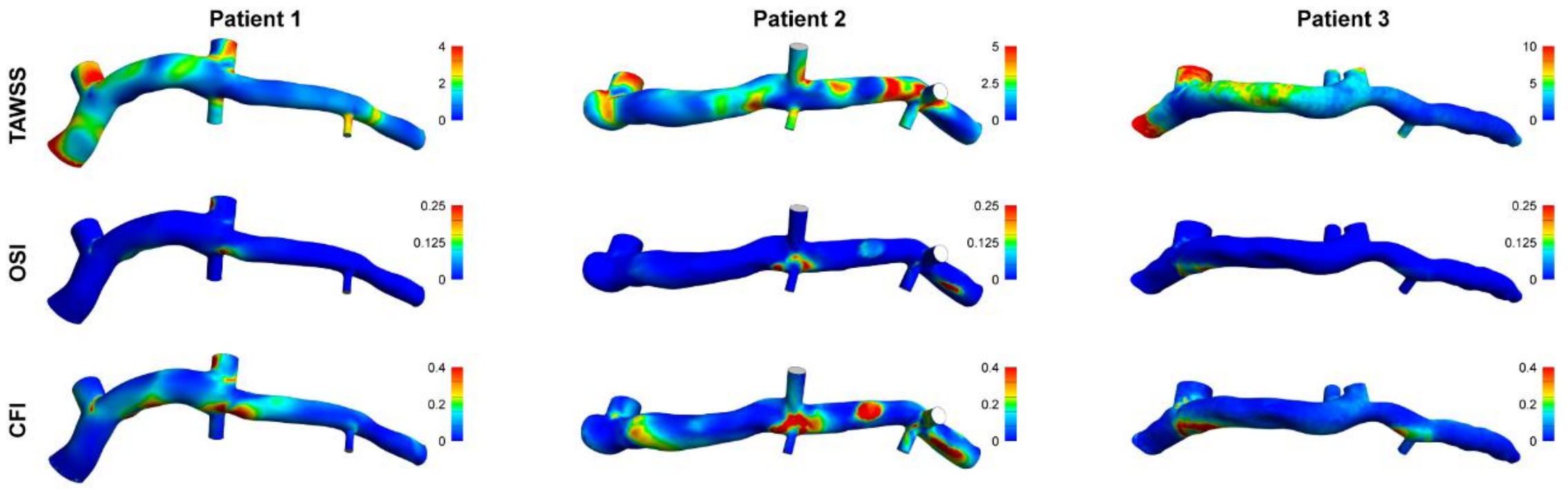

**Fig. 7.** Spatial distributions of mean time-averaged WSS (TAWSS), oscillatory shear index (OSI), and cross-flow index (CFI) for the three patient-specific coronary artery models, calculated using the fourth-order PCE emulators constructed from 505 simulations per patient.

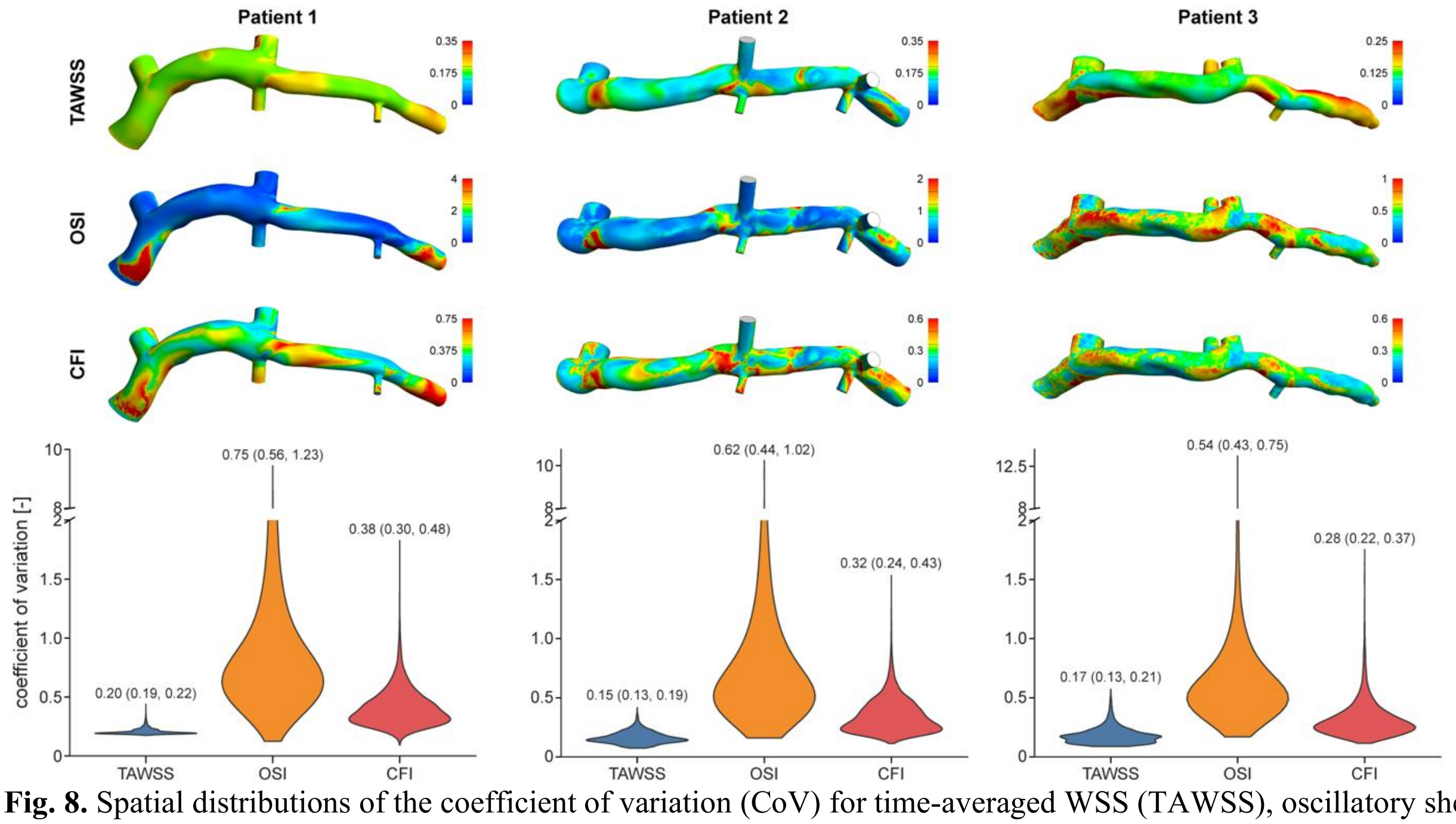

**Fig. 8.** Spatial distributions of the coefficient of variation (CoV) for time-averaged WSS (TAWSS), oscillatory shear index (OSI), and cross-flow index (CFI) for the three patient-specific coronary artery models, calculated using the fourth-order PCE emulators constructed from 505 simulations per patient. Corresponding violin plots illustrate the distribution of CoV across the vessel. Data are reported as median (Q1, Q3).

### 3.4 Total Sobol Indices and Parameter Interactions

Total Sobol indices revealed differences in the relative contributions of the uncertain parameters across QoIs and patients (Fig. 9). Overall, velocity and viscosity were the dominant sources of uncertainty, although their relative importance varied among QoIs and patients. For TAWSS, velocity exhibited the greatest sensitivity in Patient 1, with a total Sobol index of 0.57 compared with 0.43 for viscosity. Conversely, viscosity dominated in Patients 2 and 3, with total Sobol indices of 0.75 and 0.73, respectively. Density, pressure, and heart rate made negligible contributions to TAWSS in all three patients (total Sobol indices <0.015). For OSI and CFI, velocity consistently exhibited the greatest sensitivity across all patients, with total Sobol indices ranging from 0.65 to 0.87, followed by viscosity (0.19–0.38). Heart rate exhibited smaller but non-negligible contributions to OSI and CFI, with total Sobol indices ranging from 0.12 to 0.19, whereas density and pressure contributed relatively little.

The relative contributions of individual parameter effects and higher-order interactions differed among the QoIs (Fig. 9). First-order (unary) effects accounted for more than 98% of the total variance in TAWSS for all three patients, indicating that TAWSS uncertainty was primarily governed by the individual effects of velocity and viscosity, with minimal contributions from parameter interactions. In contrast, first-order effects accounted for 65–79% and 62–73% of the total variance in OSI and CFI, respectively, indicating greater contributions from interactions among uncertain parameters. Second-order (binary) interactions contributed 26%, 16%, and 19% of the total variance in OSI for Patients 1, 2, and 3, respectively, with similar contributions observed for CFI. Third-order (ternary) interactions contributed less, accounting for approximately 8% and 9% of the total variance in OSI and CFI, respectively, in Patient 1, with lower contributions observed in Patients 2 and 3. Fourth- and higher-order interactions contributed negligibly to the total variance in all patients and QoIs.

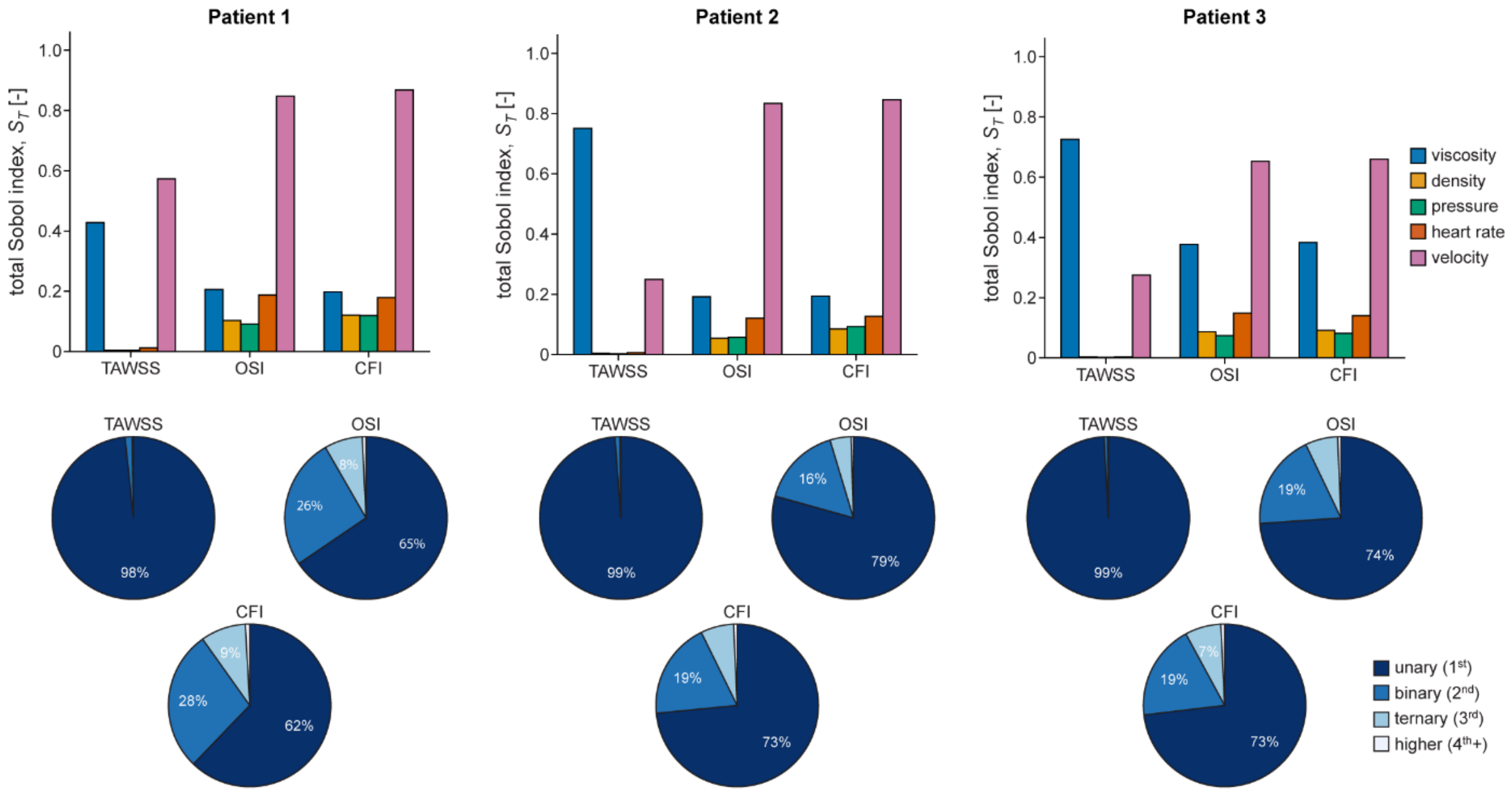

**Fig. 9.** Total Sobol sensitivity indices for the uncertain parameters influencing time-averaged WSS (TAWSS), oscillatory shear index (OSI), and cross-flow index (CFI) across the three patient-specific simulations at PCE order 4 (top row). Corresponding pie charts illustrate the relative contributions of first-order (unary), second-order (binary), third-order (ternary) and higher-order parameter interactions to the total variance in each QoI (bottom row).

The contributions of individual uncertain parameters and their specific interactions to the total variance in each QoI were further examined for each patient. For Patient 1, TAWSS variability was dominated by the individual effects of velocity (55.7%) and viscosity (41.9%), with each parameter interaction contributing less than 1% of the total variance (Fig. 10). In contrast, parameter interactions contributed substantially more to OSI and CFI variability, with the largest interactions involving velocity. The dominant second-order (binary) interactions contributing to OSI variability involved velocity paired with heart rate (7.8%), viscosity (7.3%), density (5.2%), and pressure (4.5%). Notably, several interactions involving velocity contributed more to OSI and CFI variance than the individual first-order contributions of the corresponding parameters, highlighting the importance of velocity and its interactions with other uncertain parameters in determining variability in both directional metrics. Similar trends were observed for CFI, with corresponding contributions of 7.5%, 6.7%, 6.0%, and 6.1%, respectively. Third-order (ternary) interactions made smaller individual contributions, ranging from 1.2% to 1.7% among the leading interactions for OSI and CFI.

Although similar trends were observed in Patients 2 and 3, several differences emerged relative to Patient 1 (Figs. S4 and S5). TAWSS variability was predominantly governed by uncertainty in viscosity in Patients 2 and 3 (74.4% and 72.0%, respectively), with smaller but appreciable contributions from velocity (23.9% and 26.7%, respectively). Velocity remained the largest individual contributor to OSI and CFI variability, accounting for 64.1% and 59.9% of the total variance, respectively, in Patient 2, compared with 41.5% for both QoIs in Patient 3. Notably, the individual contribution of viscosity to OSI and CFI variability was substantially greater in Patient 3 (26.0% for both QoIs) than in Patients 1 and 2. As observed in Patient 1, the dominant interactions contributing to OSI and CFI variability in Patients 2 and 3 involved velocity. Interactions between velocity and viscosity were the largest contributors to OSI variability in Patients 2 and 3, accounting for 5.5% and 6.2% of the total variance, respectively. Velocity interactions with heart rate (4.3% and 4.1%), density (2.4% and 3.6%), and pressure (2.6% and 3.1%) also contributed to OSI variability in Patients 2 and 3, respectively. Similar patterns were observed for CFI, although velocity–heart rate interactions (5.0%) slightly exceeded velocity–viscosity interactions (4.9%) in Patient 2, whereas velocity–viscosity interactions were dominant in Patient 3 (6.4%). Third-order interactions remained relatively small, with each leading interaction contributing less than 1.5% of the total variance.

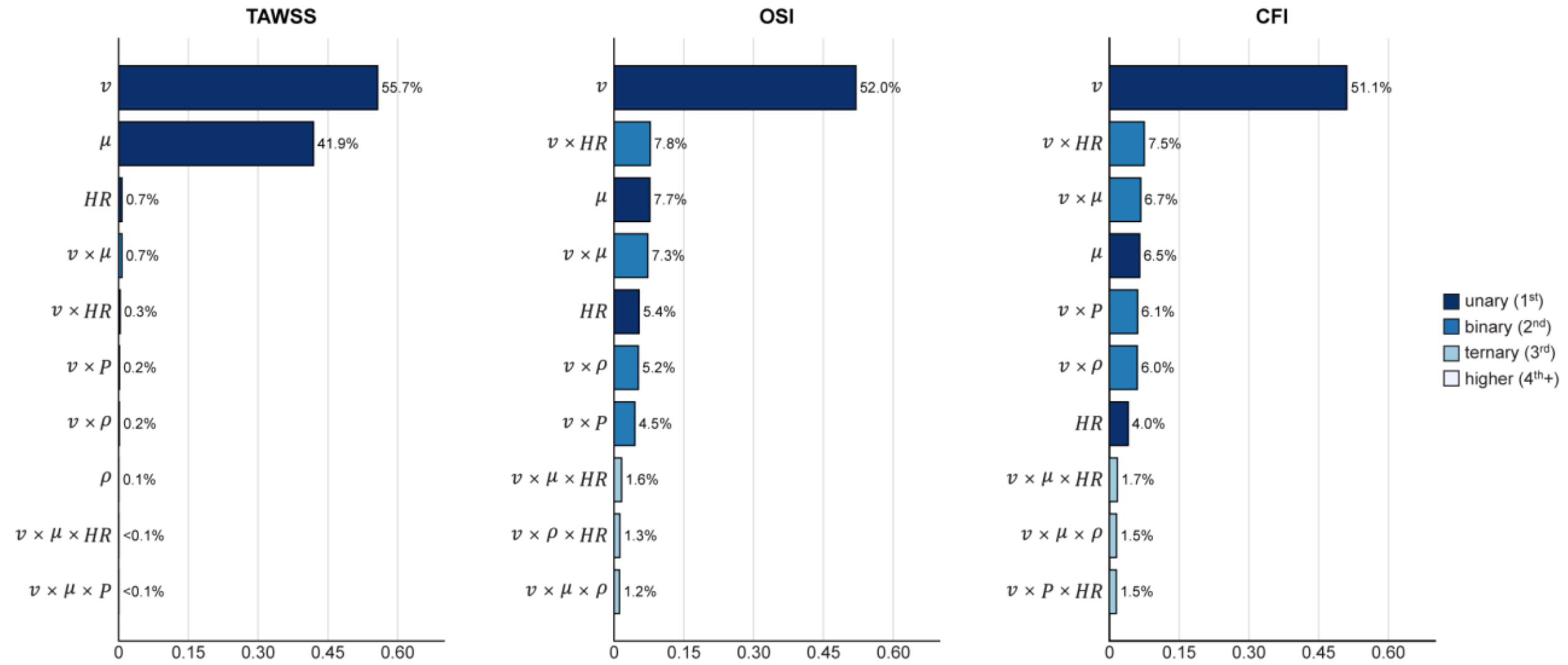

**Fig. 10.** Contributions of individual uncertain parameters and their interactions to the total variance in time-averaged WSS (TAWSS), oscillatory shear index (OSI), and cross-flow index (CFI) for Patient 1. The ten largest contributions are shown for each QoI, categorized by interaction order, with values indicating the percentage of total variance explained by each parameter or interaction.

The relative contributions of individual parameters and their interactions normalized within each interaction order further highlighted the dominant role of velocity uncertainty in determining QoI variability. In Patient 1, velocity accounted for the largest proportion of first-order (unary) effects in TAWSS (56.6%), OSI (79.4%), and CFI (82.1%), followed by viscosity (42.6%, 11.8%, and 10.5%, respectively; Fig. 11). Among second-order interactions, velocity-viscosity interactions accounted for the largest proportion of TAWSS variability (49.8%), whereas velocity-heart rate interactions were most influential for OSI (29.7%) and CFI (26.9%), followed by velocity-viscosity interactions (27.8% and 24.1%, respectively). Similarly, velocity-viscosity-heart rate interactions accounted for the largest proportions of third-order (ternary) effects for TAWSS (18.6%), OSI (21.5%), and CFI (19.0%). These trends were generally consistent in Patients 2 and 3, although viscosity dominated the first-order contributions to TAWSS (72.6% and 75.3%, respectively; Figs. S6 and S7). Velocity remained a component of the dominant second- and third-order interactions across all QoIs, with velocity-viscosity and velocity-heart rate combinations generally exhibiting the largest second-order contributions.

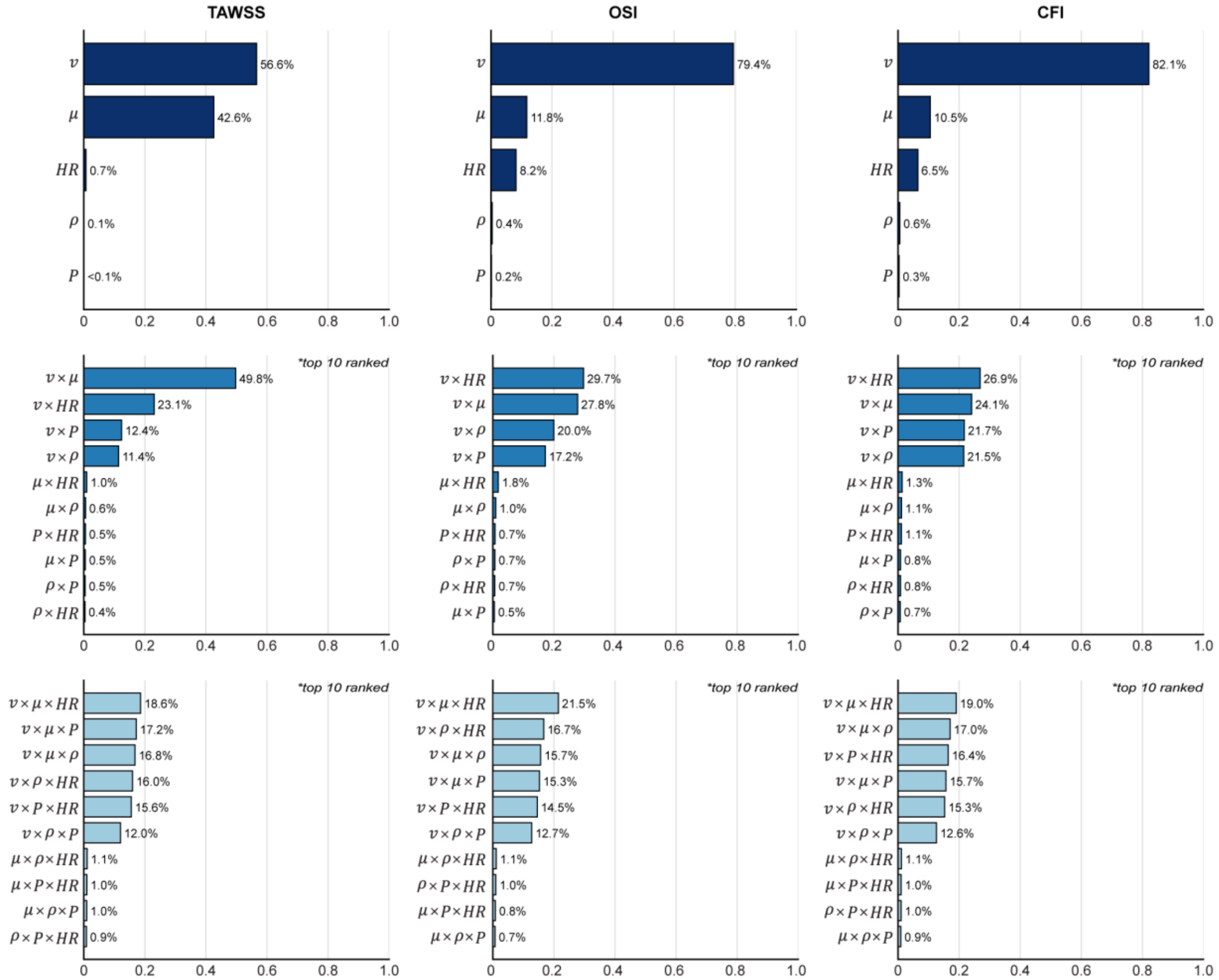

**Fig. 11.** Relative contributions of individual uncertain parameters and their second- and third-order interactions to variability in time-averaged WSS (TAWSS), oscillatory shear index (OSI), and cross-flow index (CFI) for Patient 1. First-order (top row), second-order (middle), and third-order (bottom) Sobol indices were normalized within each interaction order. The ten largest contributions are shown for second- and third-order interactions.

The relative contributions of individual uncertain parameters and their combined interactions to total QoI variance revealed consistent patterns across all three patients (Fig. 12). TAWSS variability was predominantly governed by velocity and viscosity uncertainty, with all parameter interactions collectively accounting for <2% of the total variance in each patient. In contrast, interactions accounted for substantially greater proportions of OSI (20.7–34.5%) and CFI (26.6–37.8%) variability. Velocity remained the largest individual contributor to OSI and CFI across all patients, although its contribution was lower in Patient 3 (41.5% for both QoIs) compared with Patients 1 and 2 (51.1–64.1%). Conversely, viscosity contributed more substantially to OSI and CFI variability in Patient 3 (26.0% for both QoIs) than in Patients 1 and 2 (6.5–10.4%).

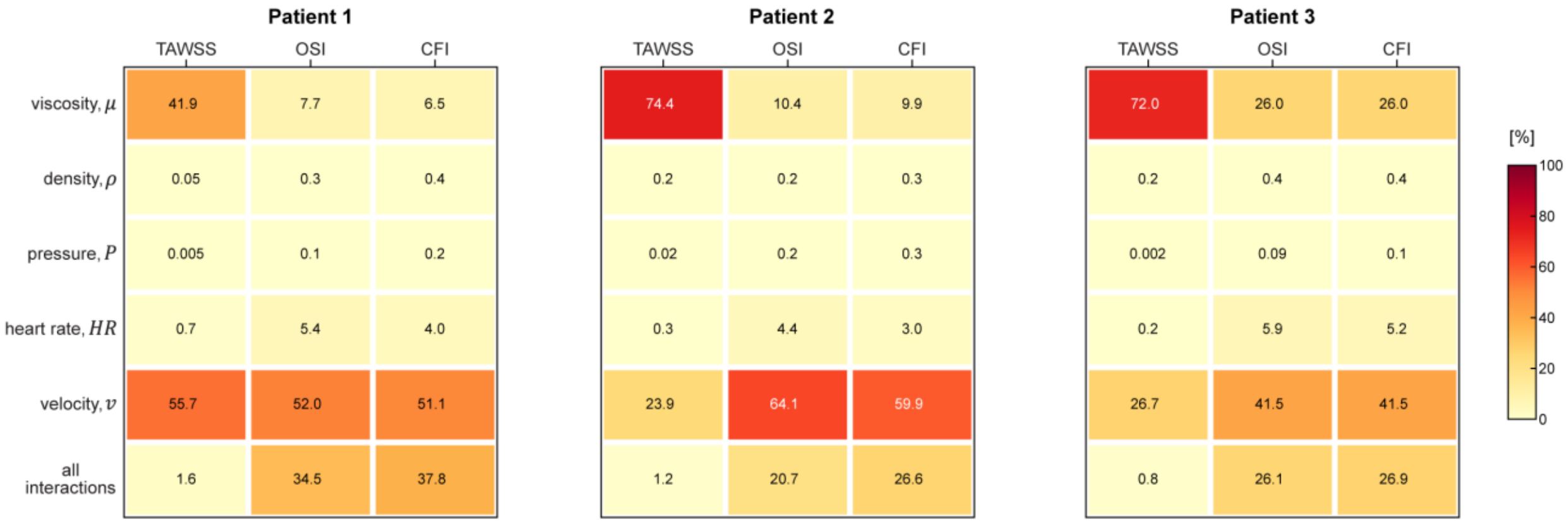

**Fig. 12.** Percentage contributions of individual uncertain parameters and their combined interactions to the total variance in time-averaged WSS (TAWSS), oscillatory shear index (OSI), and cross-flow index (CFI) for all three patients. Individual parameter contributions represent first-order Sobol indices, whereas the combined interaction contribution includes all second- and higher-order Sobol indices.

## 4. Discussion

In this study, we leveraged our previously developed UQ framework [34,35] to investigate the propagation of uncertain physiological inputs and their influence on WSS-derived QoIs in patient-specific coronary arteries under unsteady flow conditions. The framework integrates `UncertainSCI` and `FEBio` to characterize uncertainty in clinically measured velocity waveforms using PCA, together with uncertainty in pressure, viscosity, density, and heart rate. Fourth-order PCE emulators demonstrated convergence across all three patient-specific models, enabling efficient uncertainty quantification and global sensitivity analysis of TAWSS, OSI, and CFI. The results demonstrated that OSI exhibited the greatest relative uncertainty, followed by CFI and TAWSS. OSI and CFI variability was dominated by velocity waveform uncertainty, with substantial contributions from higher-order parameter interactions, particularly those involving velocity. In contrast, TAWSS variability was primarily governed by the individual effects of velocity and viscosity, with parameter interactions contributing less than 2% of the total variance. Collectively, these findings highlight the central role of velocity uncertainty in predicting coronary hemodynamics, particularly for QoIs that characterize temporal variations in WSS (e.g., OSI and CFI).

Recognizing the need to establish confidence in computational simulations to advance the diagnosis and treatment of cardiovascular disease, UQ has been increasingly integrated across cardiovascular modeling applications, including cardiac electrophysiology and electromechanics [47][48], vascular and myocardial wall mechanics [34,49–51], and medical devices [52]. In vascular hemodynamics, advances have occurred both in applying UQ to increasingly complex, patient-specific computational models and in developing efficient methodologies for uncertainty propagation and sensitivity analysis. In coronary hemodynamics, for example, UQ has been applied to patient-specific CFD simulations involving moderate CAD [35], obstructive CAD [23], coronary aneurysms associated with Kawasaki disease [53], and coronary bypass grafting [27]. Complementing these applications, multilevel and multifidelity Monte Carlo methods have demonstrated substantial reductions in computational cost by combining high-fidelity three-dimensional simulations with lower-dimensional models of healthy and diseased coronary anatomies [54]. More recently, Bayesian parameter estimation has been integrated with patient-specific myocardial perfusion measurements and multifidelity Monte Carlo methods to characterize uncertainty in coronary hemodynamic predictions while improving computational efficiency[24]. Building on these advances, the present study employed PCA to characterize physiological input uncertainty directly from repeated patient-specific invasive velocity measurements. Unlike previous studies that represented inflow waveform uncertainty using Karhunen-Loève (K-L) expansions [53], this approach captures the dominant patterns of observed cardiac cycle-to-cycle variability and enables uncertainty propagation using empirically informed distributions of waveform coefficients. Although invasive hemodynamic measurements are not routinely available, these findings highlight their value for characterizing patient-specific physiological variability and its influence on computational predictions. Future advances in non-invasive imaging and physiological measurement techniques may enable similar characterization of input uncertainty, facilitating broader and more comprehensive application of UQ in patient-specific coronary simulations.

The dominant contribution of velocity waveform uncertainty to OSI and CFI variability underscores the importance of accurate and reproducible measurements of coronary velocity waveforms. In this study, repeated invasive measurements revealed appreciable cardiac cycle-to-cycle variability, with relative RMSEs of 13.4%, 13.9%, and 7.4% for Patients 1, 2, and 3, respectively (Fig. 4). PCA further revealed differences in dimensionality of waveform variability, with three components capturing more than 95% of the variance in Patient 3, compared with four components required to approach or exceed that threshold in Patients 1 and 2 (Fig. 3). This variability likely reflects a combination of physiological fluctuations and measurement-related uncertainty (i.e., noise), highlighting the importance of distinguishing these sources when characterizing uncertainty. Notably, Patient 3, who exhibited the lowest cycle-to-cycle velocity variability, also demonstrated lower Sobol sensitivity indices for velocity waveform uncertainty in OSI and CFI (Figs. 9 and 12). This observation suggests that differences in measured waveform variability may influence the relative contributions of velocity uncertainty to predicted hemodynamic QoIs, further emphasizing the importance of accurately characterizing patient-specific physiological inputs. Moreover, improved measurement quality may reduce noise-related variability and, consequently, decrease the effective dimensionality of the uncertain input space and computational costs of the UQ analyses.

The sensitivity analysis for TAWSS demonstrated trends consistent with our previous investigation of patient-specific coronary CFD simulations under steady flow conditions [35], which utilized data from Patient 1 in the present study. In both analyses, uncertainty in blood viscosity and velocity emerged as the dominant contributors to variability in WSS magnitude. However, steady-flow simulations cannot capture the temporal variations in WSS required to quantify OSI and CFI. By extending the UQ framework to unsteady flow conditions, the present study revealed distinct sensitivity characteristics for these hemodynamic QoIs, including a greater influence of velocity waveform uncertainty and higher-order parameter interactions. These findings underscore the importance of incorporating pulsatile flow and evaluating multiple WSS-derived quantities to comprehensively characterize uncertainty in patient-specific coronary hemodynamic predictions.

Ultimately, the clinical significance of UQ in patient-specific coronary CFD modeling depends on whether uncertainty in predicted hemodynamics QoIs influences the interpretation of disease progression or clinical decision-making. Previous studies have demonstrated associations between WSS-derived quantities and coronary plaque progression, changes in plaque composition, and vulnerability [5,6,14–16,55]. Importantly, TAWSS, OSI, and CFI characterize distinct aspects of the coronary hemodynamic environment, suggesting that uncertainty in these quantities may have different implications for assessing disease risk. Although the present study demonstrates appreciable variability in these hemodynamic predictions, whether this uncertainty translates into meaningful differences in the identification of high-risk coronary regions or subsequent clinical decisions remains unknown. Consequently, future studies should investigate whether uncertainty in WSS-derived metrics alters the spatial identification of coronary segments exposed to atheroprone hemodynamics conditions and their associated with subsequent plaque progression. Establishing this relationship through longitudinal clinical studies will be essential for determining the clinical value of incorporating UQ into patient-specific coronary simulations.

Implementing UQ in biomechanical simulations requires careful consideration of computational efficiency and data-management demands. To address these computational demands, nonintrusive stochastic collocation methods and multilevel, multifidelity Monte Carlo approaches have demonstrated substantial computational savings in cardiovascular simulations through efficient sampling strategies and the use of lower-fidelity models [22,54]. As employed in the present study, WAFP sampling coupled with PCE emulators provides an efficient approach for uncertainty propagation and global sensitivity analysis with fewer model evaluations than conventional Monte Carlo methods [35,41]. These efficiencies are particularly important for unsteady CFD simulations and more complex biomechanical models involving nonlinear material behavior, large deformations, and coupled multiphysics phenomena. However, computational efficiency must be balanced against the accuracy of the UQ estimates, requiring appropriate convergence assessments and emulator validation. Moreover, the substantial volume of data generated by these simulations presents additional challenges for data storage, management, and postprocessing. For reference, the 3,770 simulations performed in this study generated >21 TB of data, highlighting the considerable data management demands associated with ensemble-based UQ analyses. Consequently, automated workflows for simulation execution, data management, and data post-processing are essential for ensuring consistency, traceability, and reproducibility. Continued advancements in efficient sampling strategies, high-performance computing, and automated data processing will facilitate the application of UQ to larger patient cohorts and increasingly complex biomechanical models.

Although this study advances the integration of UQ into patient-specific coronary hemodynamic simulations under pulsatile flow conditions, several limitations must be acknowledged. The modeling assumptions were primarily intended to reduce computational cost and facilitate interpretation of the individual and interaction effects of uncertain inputs. First, the coronary arteries were modeled with rigid walls, thereby neglecting vessel deformation and uncertainty in wall thickness and material properties. Although fluid–structure interaction methods are well established [45], incorporating vessel deformation and anisotropic material behavior would increase computational costs and expand the dimensionality of the UQ analysis by introducing additional uncertain parameters [26,27,34]. Second, blood was modeled as an incompressible Newtonian fluid with an uncertain but assumed constant viscosity. Although this assumption has been shown to yield WSS distributions comparable to those obtained using shear-thinning models in patient-specific coronary CFD simulations [56], non-Newtonian effects may remain relevant in regions of disturbed flow and could influence predicted OSI and CFI values and their corresponding sensitivity indices. Third, the patient-specific geometries were treated as deterministic, neglecting uncertainties associated with image resolution, image noise, and lumen segmentation. Although geometric uncertainty can influence predicted hemodynamics [30,31], incorporating this additional source of uncertainty presents computational and methodological challenges, requiring automated generation, discretization, and analysis of multiple geometric realizations. Deep-learning-based approaches, such as the geometric UQ framework coupled with CFD developed by Maher et al. [31], offer a promising strategy for characterizing segmentation uncertainty and incorporating geometric variability into future UQ analyses. Finally, physiological input uncertainty was characterized using repeated invasive measurements from three patients, with velocity waveform variability estimated from a limited number of cardiac cycles. Although these measurements enabled patient-specific characterization of observed cycle-to-cycle variability, they may not capture the full range of physiological fluctuations occurring over longer measurement periods or under different physiological conditions. Future studies involving larger patient cohorts and more extensive physiological measurements will be important for assessing the generalizability of the uncertainty distributions and their influence on predicted hemodynamic QoIs.

## 5. Conclusion

In conclusion, this study characterized and incorporated uncertainty in hemodynamic measurements used as inputs to patient-specific, unsteady coronary CFD simulations. The results demonstrated that uncertainty in physiological inputs propagates into appreciable variability in WSS-derived quantities, with OSI and CFI exhibiting greater relative uncertainty than TAWSS.

Velocity waveform uncertainty emerged as the dominant contributor to OSI and CFI variability, with higher-order parameter interactions also playing an important role. These findings highlight the importance of accounting for physiological uncertainty when interpreting patient-specific coronary hemodynamic predictions. Future efforts will focus on incorporating additional sources of uncertainty, while leveraging advances in medical imaging and physiological measurement technologies to improve the characterization of patient-specific inputs and reduce associated uncertainties. Ultimately, integrating comprehensive UQ with improved data acquisition will strengthen the credibility of computational simulations and confidence in their predictions as these approaches are translated into the clinical practice to advance diagnosis, prognosis, and treatment optimization.

## CRediT authorship contribution statement

**Muhammad Usman:** Conceptualization, Methodology, Formal analysis, Writing - Original Draft. **Nafis Arefin:** Formal analysis, Writing - Original Draft. **Akil Narayan:** Methodology, Software, Formal analysis, Writing - Review & Editing. **Lucas H. Timmins:** Conceptualization, Methodology, Formal analysis, Writing - Original Draft, Supervision

## Acknowledgments

This research was funded, in part, by the National Institutes of Health - award numbers R01 HL150608 (L.H.T.), U24 EB029012 (A.N.). CFD simulations were conducted at the Texas A&M University High Performance Research Computing (HPRC) facility. The authors gratefully acknowledge Prof. Gerard Ateshian at Columbia University for his valuable technical guidance and assistance with FEBio. The authors would also like to extend their acknowledgment to Drs. Habib Samady and David Molony for providing the clinical data.

## Supplementary Material

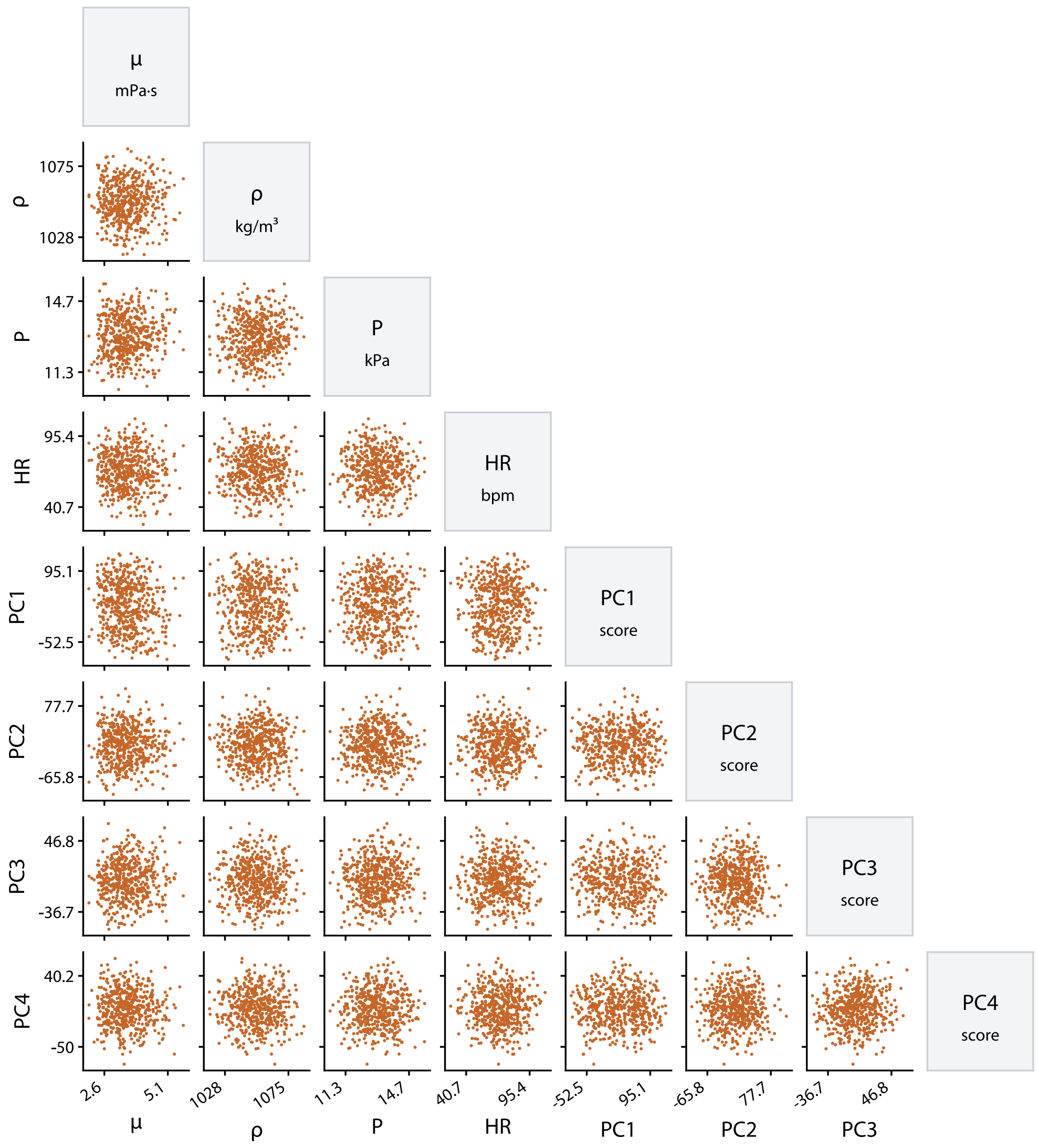

**Fig. S1**: Parameter distributions at order 4 (505 parameter sets). The velocity waveform was described by principal component (PC) scores 1-4.

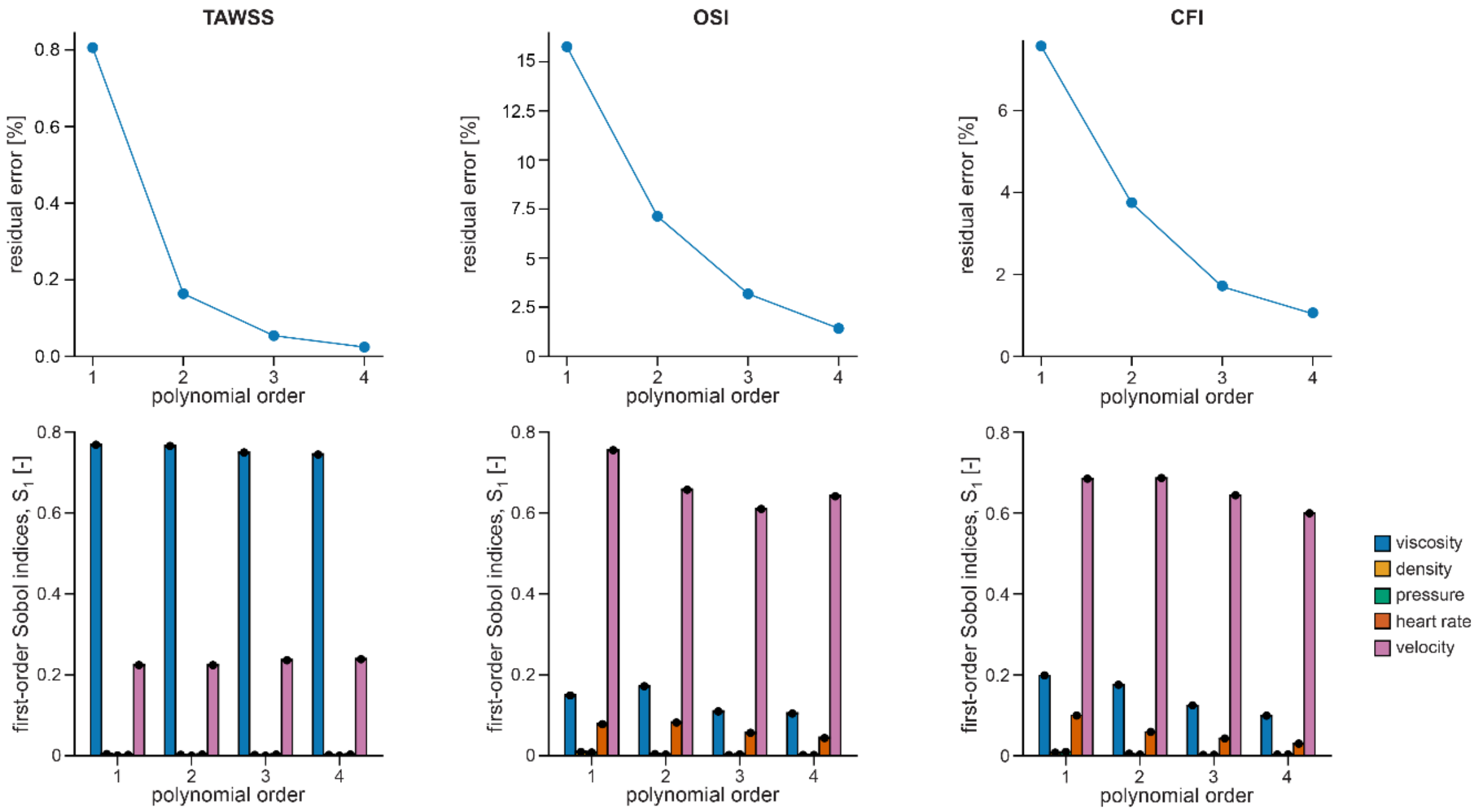

**Fig. S2**: PCE convergence analysis for Patient 2 across PCE orders 1-4. Spatially averaged residual errors (top row) and first-order Sobol indices (bottom row) are presented for time-averaged WSS (TAWSS), oscillatory shear index (OSI), and cross-flow index (CFI). Results are shown for three independent PCE runs at each order, with individual runs indicated by black markers. Data are reported as mean ± standard deviation.

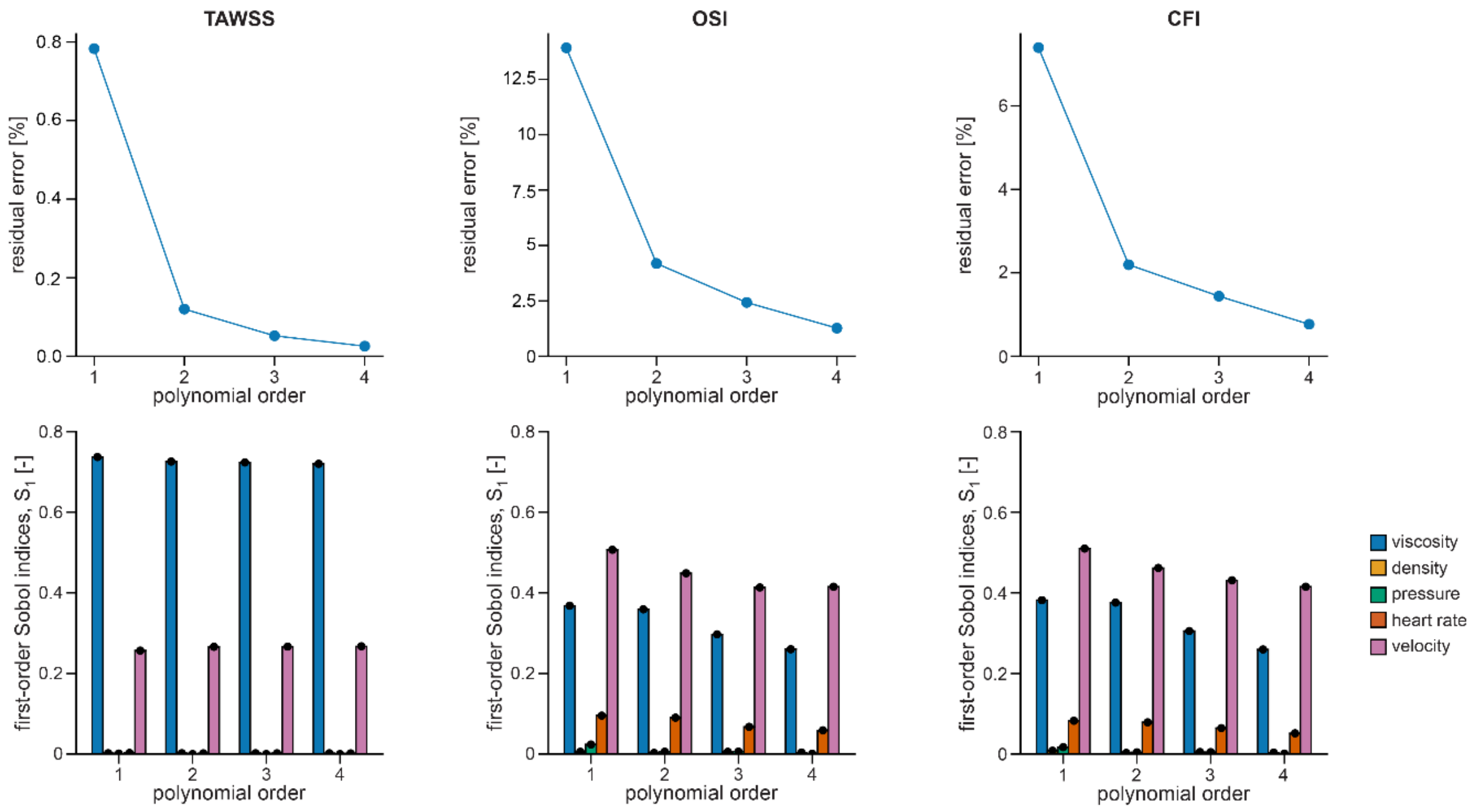

**Fig. S3**: PCE convergence analysis for Patient 3 across PCE orders 1-4. Spatially averaged residual errors (top row) and first-order Sobol indices (bottom row) are presented for time-averaged WSS (TAWSS), oscillatory shear index (OSI), and cross-flow index (CFI). Results are shown for three independent PCE runs at each order, with individual runs indicated by black markers. Data are reported as mean ± standard deviation.

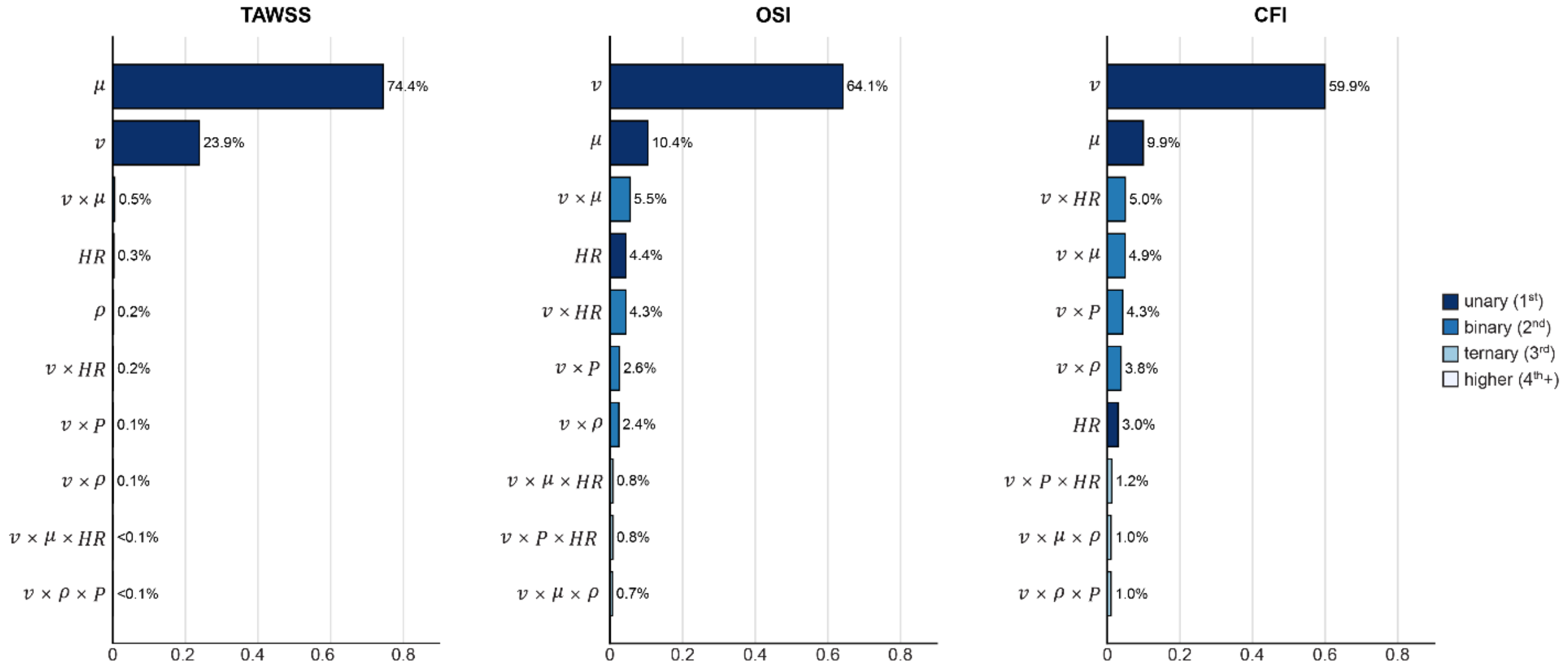

**Fig. S4**: Contributions of individual uncertain parameters and their interactions to the total variance in time-averaged WSS (TAWSS), oscillatory shear index (OSI), and cross-flow index (CFI) for Patient 2. The ten largest contributions are shown for each QoI, categorized by interaction order, with values indicating the percentage of total variance explained by each parameter or interaction.

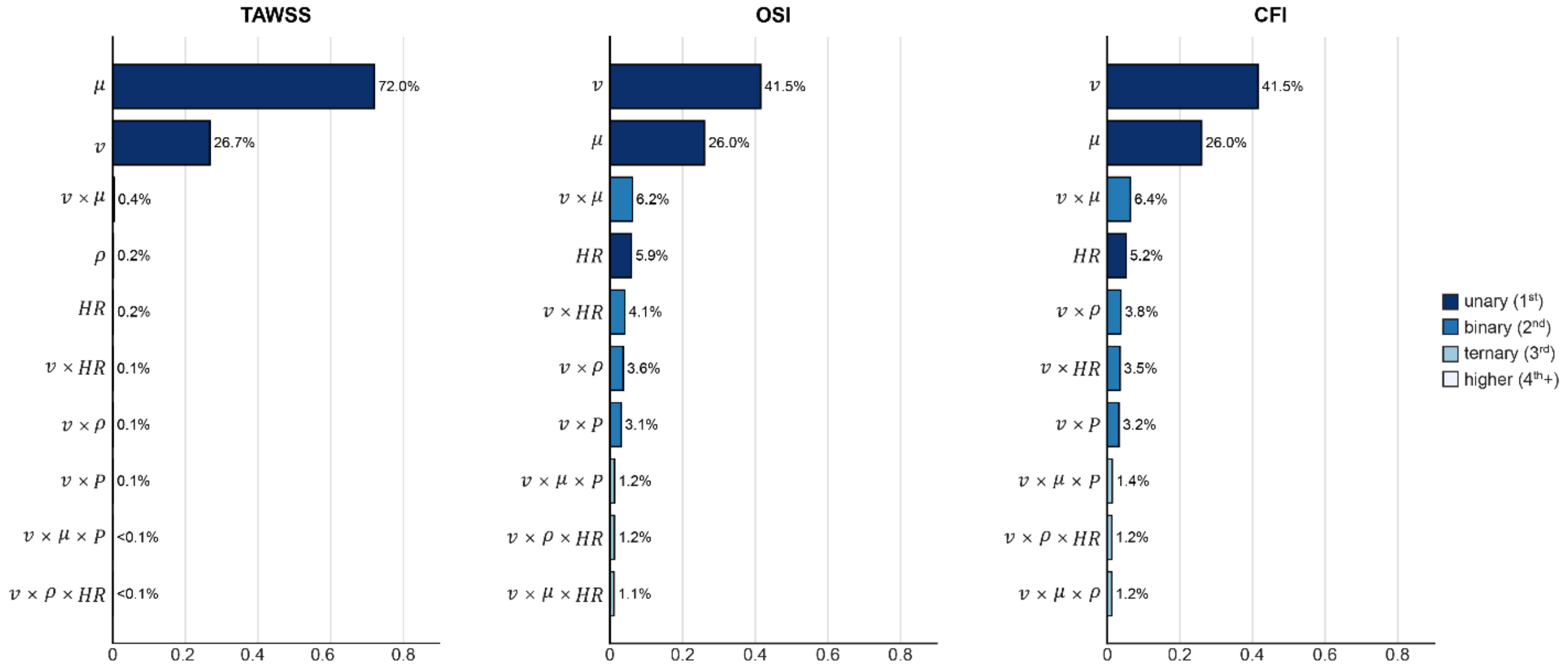

**Fig. S5**: Contributions of individual uncertain parameters and their interactions to the total variance in time-averaged WSS (TAWSS), oscillatory shear index (OSI), and cross-flow index (CFI) for Patient 3. The ten largest contributions are shown for each QoI, categorized by interaction order, with values indicating the percentage of total variance explained by each parameter or interaction.

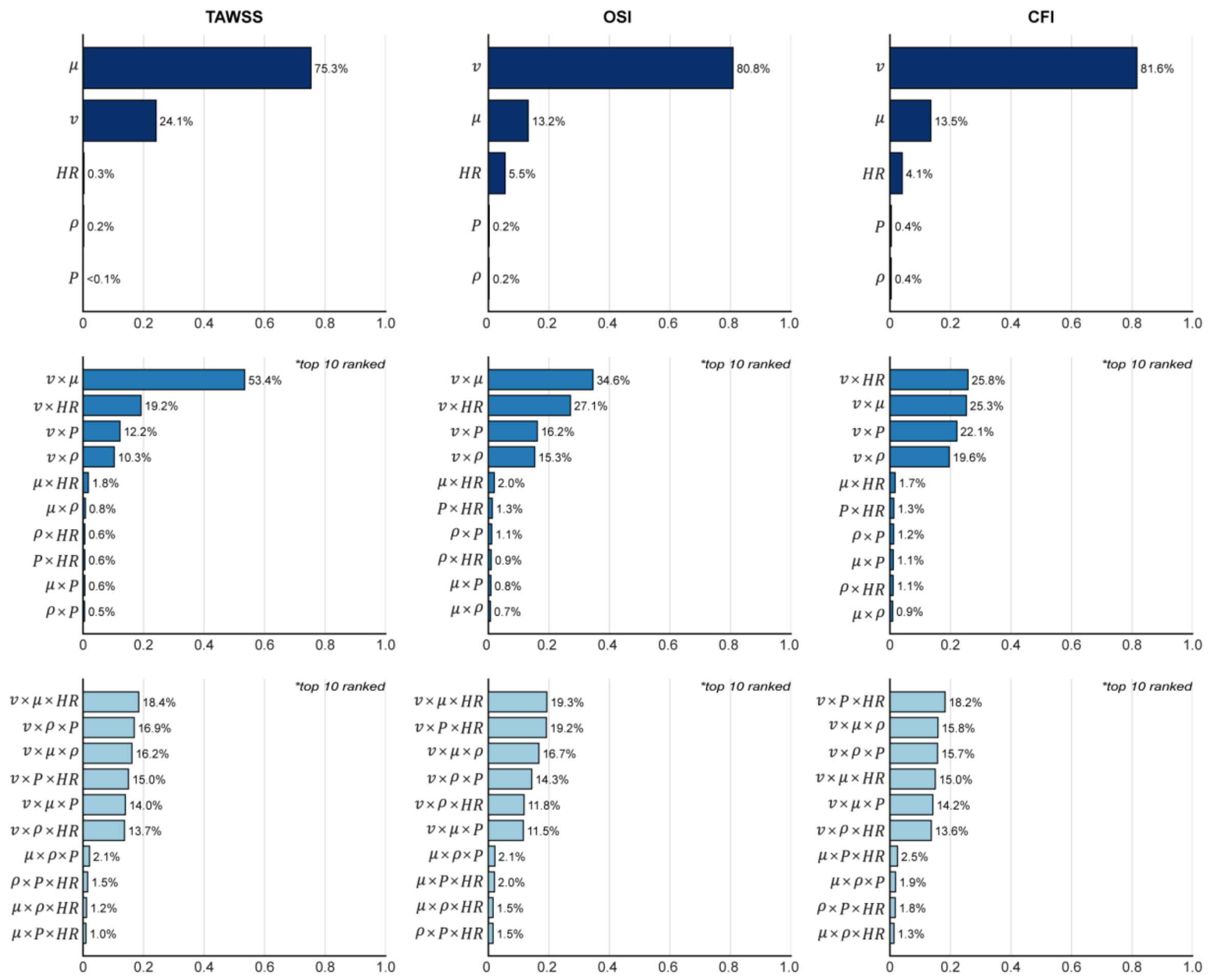

**Fig. S6**: Relative contributions of individual uncertain parameters and their second- and third-order interactions to variability in time-averaged WSS (TAWSS), oscillatory shear index (OSI), and cross-flow index (CFI) for Patient 2. First-order (top row), second-order (middle), and third-order (bottom) Sobol indices were normalized within each interaction order. The ten largest contributions are shown for second- and third-order interactions.

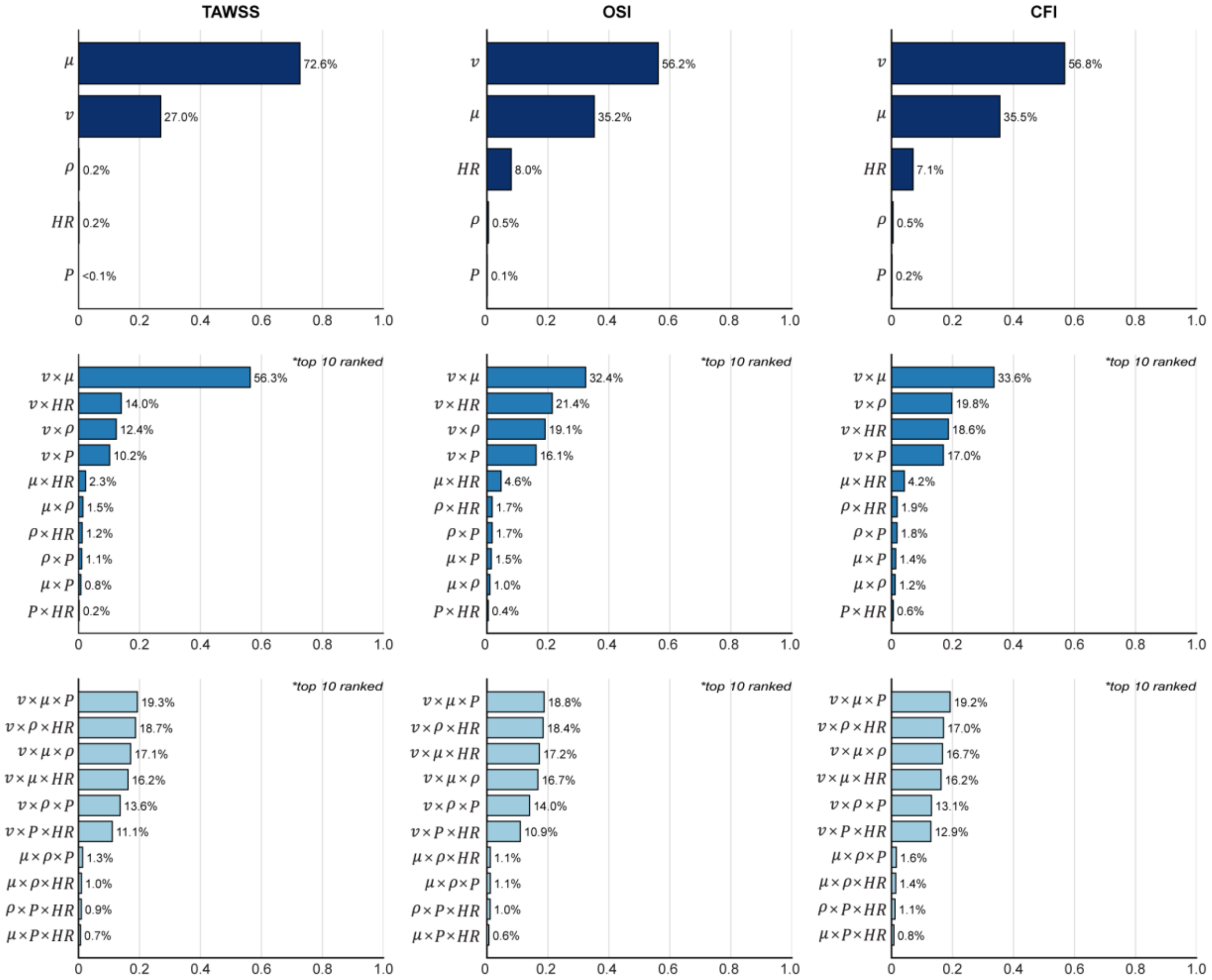

**Fig. S7**: Relative contributions of individual uncertain parameters and their second- and third-order interactions to variability in time-averaged WSS (TAWSS), oscillatory shear index (OSI), and cross-flow index (CFI) for Patient 3. First-order (top row), second-order (middle), and third-order (bottom) Sobol indices were normalized within each interaction order. The ten largest contributions are shown for second- and third-order interactions.